\documentclass[twocolumn]{aastex631}

\usepackage{amsmath, amssymb}
\usepackage{xspace}
\usepackage[super]{nth} 
\usepackage{multirow}
\usepackage[T1]{fontenc}

\newcommand{\kepler}{\emph{Kepler}\xspace}
\newcommand{\gaia}{\emph{Gaia}\xspace}
\defcitealias{DressingCharbonneau2015}{DC15}
\usepackage[normalem]{ulem}

\accepted{for publication in The Astronomical Journal; October 15th, 2023}

\shorttitle{Earth-sized Habitable Zone Planets around \emph{Kepler}'s M Dwarfs}
\shortauthors{Bergsten et al.}

\begin{document}

\title{No Evidence for More Earth-sized Planets in the Habitable Zone of \emph{Kepler}'s M versus FGK Stars}

\correspondingauthor{Galen J. Bergsten}
\email{gbergsten@arizona.edu}

\author[0000-0003-4500-8850]{Galen J. Bergsten}
\affil{Lunar and Planetary Laboratory, The University of Arizona, Tucson, AZ 85721, USA}

\author[0000-0001-7962-1683]{Ilaria Pascucci}
\affil{Lunar and Planetary Laboratory, The University of Arizona, Tucson, AZ 85721, USA}

\author[0000-0003-3702-0382]{Kevin K. Hardegree-Ullman}
\affil{Steward Observatory, The University of Arizona, Tucson, AZ 85721, USA}

\author[0000-0002-3853-7327]{Rachel B. Fernandes}
\altaffiliation{President's Postdoctoral Fellow}
\affil{Department of Astronomy \& Astrophysics, Center for Exoplanets and Habitable Worlds, The Pennsylvania State University, University Park, PA 16802, USA}
\affil{Center for Exoplanets and Habitable Worlds, 525 Davey Laboratory, The Pennsylvania State University, University Park, PA, 16802, USA}
\affil{Lunar and Planetary Laboratory, The University of Arizona, Tucson, AZ 85721, USA}

\author[0000-0002-8035-4778]{Jessie L. Christiansen}
\affiliation{NASA Exoplanet Science Institute, IPAC, MS 100-22, Caltech, 1200 E. California Blvd, Pasadena, CA 91125}

\author[0000-0002-1078-9493]{Gijs D. Mulders}
\affil{Facultad de Ingenier\'ia y Ciencias, Universidad Adolfo Ib\'a\~nez, Av.\ Diagonal las Torres 2640, Pe\~nalol\'en, Santiago, Chile}
\affil{Millennium Institute for Astrophysics, Chile}


\newcommand {\galen}[1]  {\textit{\textcolor{magenta}{[Galen: #1]}}}
\newcommand {\ilaria}[1]  {\textit{\textcolor{violet}{[IP: #1]}}}
\newcommand {\new}[1]  {\textbf{\textcolor{magenta}{#1}}}
 
\begin{abstract}
Reliable detections of Earth-sized planets in the habitable zone remain elusive in the \emph{Kepler} sample, even for M dwarfs. The \emph{Kepler} sample was once thought to contain a considerable number of M dwarf stars ($T_\mathrm{eff} < 4000$\,K), which hosted enough Earth-sized ($[0.5,1.5]$\,R$_\oplus$) planets to estimate their occurrence rate ($\eta_\oplus$) in the habitable zone. However, updated stellar properties from \emph{Gaia} have shifted many \emph{Kepler} stars to earlier spectral type classifications, with most stars (and their planets) now measured to be larger and hotter than previously believed. Today, only one partially-reliable Earth-sized candidate remains in the optimistic habitable zone, and zero in the conservative zone.
Here we performed a new investigation of \emph{Kepler}'s Earth-sized planets orbiting M dwarf stars, using occurrence rate models with considerations of updated parameters and candidate reliability. Extrapolating our models to low instellations, we found an occurrence rate of $\eta_\oplus={8.58}_{-8.22}^{+17.94}\%$ for the conservative habitable zone (and ${14.22}_{-12.71}^{+24.96}\%$ for the optimistic), consistent with previous works when considering the large uncertainties. Comparing these estimates to those from similarly comprehensive studies of Sun-like stars, we found that the current \emph{Kepler} sample does not offer evidence to support an increase in $\eta_\oplus$ from FGK to M stars. While the \emph{Kepler} sample is too sparse to resolve an occurrence trend between early and mid-to-late M dwarfs for Earth-sized planets, studies including larger planets and/or data from the \emph{K2} and \emph{TESS} missions are well-suited to this task.
\end{abstract}


\section{Introduction} \label{sec:intro}

In the ongoing search for life in the Universe, one standing question is the occurrence rate (or frequency) of Earth-sized planets in the habitable zone. Studies to this point are often focused on Sun-like (FGK) stars, in part because they are most reflective of our own solar system and thus life as we know it. Sharing this focus, the \kepler{} mission \citep{Borucki2010, Borucki2016} sought to detect Earth-sized habitable zone planets, with long-term monitoring preferentially targeting Sun-like stars. Mechanical failures rendered the \kepler{} telescope unable to continue viewing its target field before it could provide sufficiently numerous confident detections of Earth-sized planets at $\sim1$\,year orbital periods to enable robust occurrence calculations. As such, many estimates for the occurrence rate of Earth-sized habitable zone planets ($\eta_\oplus$) around FGK stars are based on extrapolations from the close-in population where \kepler{}'s survey completeness for small planets is comparably higher than in the habitable zone. Estimates for $\eta_\oplus$ around Sun-like stars span from $\sim1-100\%$ depending on various physical and statistical considerations, with recent estimates tending towards $10-40\%$ (see e.g., discussions in \citealp{KunimotoMatthews2020,Bergsten2022}).

The habitable zone for M dwarfs occurs at shorter orbital periods ($\sim30-140$\,days), meaning Earth-sized habitable zone planets around these stars would have ideally been detected within \kepler{}'s lifespan. Yet the survey's selection of fewer M dwarfs for monitoring, coupled with its generally poor sensitivity to fainter stars, meant that the \kepler{} population contained only a few detections of these planet candidates around M dwarfs. Nonetheless, \citet{DressingCharbonneau2015} leveraged the available \kepler{} information to identify a sample of $156$ planets around M dwarf ($T_\mathrm{eff} < 4000$\,K, $\log{g} > 3$) stars, and estimated (via the inverse detection efficiency method) the occurrence rate of Earth-sized habitable zone planets. For $[1.0,1.5]\,R_\oplus$ planets, they estimated an occurrence of $\eta_\oplus = 15.82^{+16.60}_{-6.54}\%$ for a conservative habitable zone (between the moist and maximum greenhouse boundaries, \citealp{Kopparapu2013}) and $24.28^{+17.58}_{-8.39}\%$ for an optimistic habitable zone (between the recent Venus and early Mars boundaries). Separately, \citet{Mulders2015b} used the \kepler{} Q1-Q16 planet candidate sample \citep{Mullally2015} to find that M dwarfs have $\sim3.5$ times more small planets than FGK stars. While the \citet{Mulders2015b} estimate was derived from the close-in ($P<50$\,days) population, it has since been applied to scale $\eta_\oplus$ estimates from FGK stars as an approximation for Earth-sized habitable zone planets around M dwarfs (see e.g., \citealp{KHU2023}).


Several recent advancements have since changed the field of \kepler{} demographics, including the release of the final \kepler{} \texttt{DR25} catalog \citep{Thompson2018} and stellar parameter revisions informed by the \gaia{} mission (\citealp{GaiaDR2}, e.g., \citealp{Berger2020}). The work of \citet{Hsu2020} incorporated these advancements by using \texttt{DR25} in conjunction with information from \gaia{} \texttt{DR2} and the 2MASS point source catalogue \citep{Skrutskie2006}. They estimated an occurrence rate of $33^{+10}_{-12}\%$ for $[0.75,1.5]\,R_\oplus$ planets in the conservative habitable zone around M dwarfs, and found that $\eta_\oplus$ estimates are sensitive to the choice of modeling priors in this regime of sparse detections. \citet{Hsu2020} also found that the small planet occurrence rates around M dwarfs are comparable to those of FGK stars \citep{Hsu2019} when evaluated at similar instellations, thus raising concerns that the factor of $3.5$ offset between M and FGK stars might not be applicable to the habitable zone.

A third advancement which merits consideration are new statistical treatments of candidate reliability \citep{Bryson2020-OG-Reliability, Bryson2020}, which serves to better account for false positives and false alarms. To date, no study has incorporated \kepler{} candidate reliability alongside \texttt{DR25} and \gaia{}-revised parameters to revisit the question of Earth-sized habitable zone planets around \kepler{}'s M dwarfs. Here we seek to do just that, providing an updated look into the occurrence rate of such planets around the most abundant type of star in the Galaxy.

In Section~(\ref{sec:DC15}), we outline the motivation for a new study by investigating how previous works and planet samples have been impacted by recent advancements in the field. We describe the current \kepler{} sample of Earth-sized planets around M dwarfs in Section~(\ref{sec:ModernSample}), and introduce our approaches to calculating planet occurrence and fitting population models in Section~(\ref{sec:Methods}). We discuss our results in Section~(\ref{sec:Results}) and describe the habitable zone occurrence rates in Section~(\ref{sec:HZ}). In Section~(\ref{sec:FGK}), we provide an updated comparison of Earth-sized planet occurrence rates within and across spectral types, before summarizing our work in Section~(\ref{sec:Conclusions}).

\section{Need for an Updated Investigation} \label{sec:DC15}
Before performing an updated analysis with the current \kepler{} sample, we examined how the properties of planets originally studied in \citet{DressingCharbonneau2015} (hereafter \citetalias{DressingCharbonneau2015}) had been impacted by subsequent studies. Of the 156 planets presented in \citetalias{DressingCharbonneau2015} (their Table 9), there were 145 planets\footnote{The candidate planet K04427.01 appeared with $1.56_{-0.23}^{+0.25}\,R_\oplus$ and $0.17_{-0.05}^{+0.06}\,I_\oplus$ in \citetalias{DressingCharbonneau2015}, which fell outside of the original $0.2\,I_\oplus$ lower bound. After updating properties with \gaia{}, the host star has $T_\mathrm{eff}=3895\,K$, and the candidate has a habitable zone instellation of $0.32\pm0.03\,I_\oplus$. However, the planet radius of $1.79_{-0.09}^{+0.12}\,R_\oplus$ exceeds the ``Earth-sized" definition at $1\sigma$, so it is excluded from this discussion (but is considered for our analysis of the current \kepler{} sample; Section~\ref{sec:ModernSample}).} around 85 stars with planet parameters within $[0.5,4]\,R_\oplus$ and $[0.2,400]\,I_\oplus$, where $I_\oplus$ denotes the instellation flux of (present-day) Earth. Cross-matching with the \kepler{} \texttt{DR25} catalog \citep{Thompson2018}, we found that five planet candidates are now considered false positives according to automated classification via Robovetter \citep{Coughlin2017}. We include a brief discussion of objects with contrasting dispositions between \texttt{DR25} and subsequent literature in Appendix~(\ref{app:FPs}).

For the remaining 140 planets around 83 stars, we cross-matched host stars with the \gaia{}-\kepler{} Stellar Properties Catalog \citep{Berger2020}. We found that seven stars (hosting fourteen planets) were not included in the updated catalog (typically because of low-quality photometry or lacking a \gaia{} \texttt{DR2} parallax), and we thus excluded these objects from the following analysis. This included one of six planets which, within the original $1\sigma$ uncertainties, could have fallen within the \citetalias{DressingCharbonneau2015} bounds for an Earth-sized habitable zone planet ($[1,1.5]\,R_\oplus$, $[0.23,1.54]\,I_\oplus$) for an optimistic habitable zone. This planet was detected in \citetalias{DressingCharbonneau2015} with a radius of $1.03$\,R$_\oplus$ and subsequently appeared as K01681.04 in \kepler{} \texttt{DR24} ($0.77$\,R$_\oplus$, \citealp{Coughlin2016}). However, K01681.04 later appeared in \kepler{} \texttt{DR25} with a radius of $10.39$\,R$_\oplus$, so it is unlikely that this planet would now contribute to $\eta_\oplus$ (barring an uncharacteristic order of magnitude decrease in stellar radius with \gaia{}).

From the remaining 126 planets orbiting 76 stars, we used the \gaia{}-updated stellar mass ($M_*$) estimates from \citet{Berger2020} along with \texttt{DR25} orbital periods ($P$) to calculate planet semi-major axes ($a$) with Kepler's law for a negligible planet mass:
\begin{equation}\label{eqn:KeplersLaw}
    a = \left[\frac{G M_* P^2}{4 \pi^2}\right]^{1/3},
\end{equation}
and used these in conjunction with \citet{Berger2020} luminosities ($L_*$) to estimate updated planetary instellations ($I$):
\begin{equation}\label{eqn:Flux}
    I = \frac{L_*}{4\pi a^2}.
\end{equation}
We also updated planet radii by multiplying the \texttt{DR25} planet-to-star radius ratios with the \citet{Berger2020} stellar radii. For all above calculations, we propagated uncertainties with Monte Carlo sampling of split-normal distributions using the $1\sigma$ uncertainties provided in \citet{Berger2020} for stellar parameters and in \texttt{DR25} for planetary parameters.\footnote{Our planet parameters are functionally identical to those of \citet{Berger2020b}, who use a similar methodology and set of input parameters, with only small ($<1\%$) variations attributable to our Monte Carlo sampling.}

\begin{figure*}[ht!]
     \centering\begin{minipage}[c]{0.6\textwidth}
       \includegraphics[width=\textwidth]{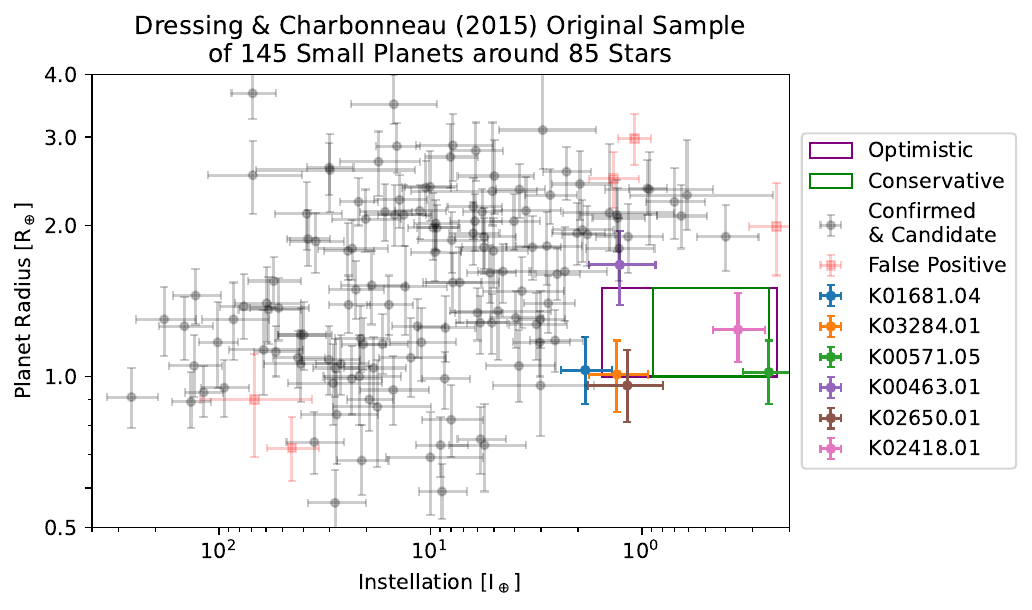}
    \end{minipage}\hfill\\
    \begin{minipage}[c]{\textwidth}
        \includegraphics[width=\textwidth]{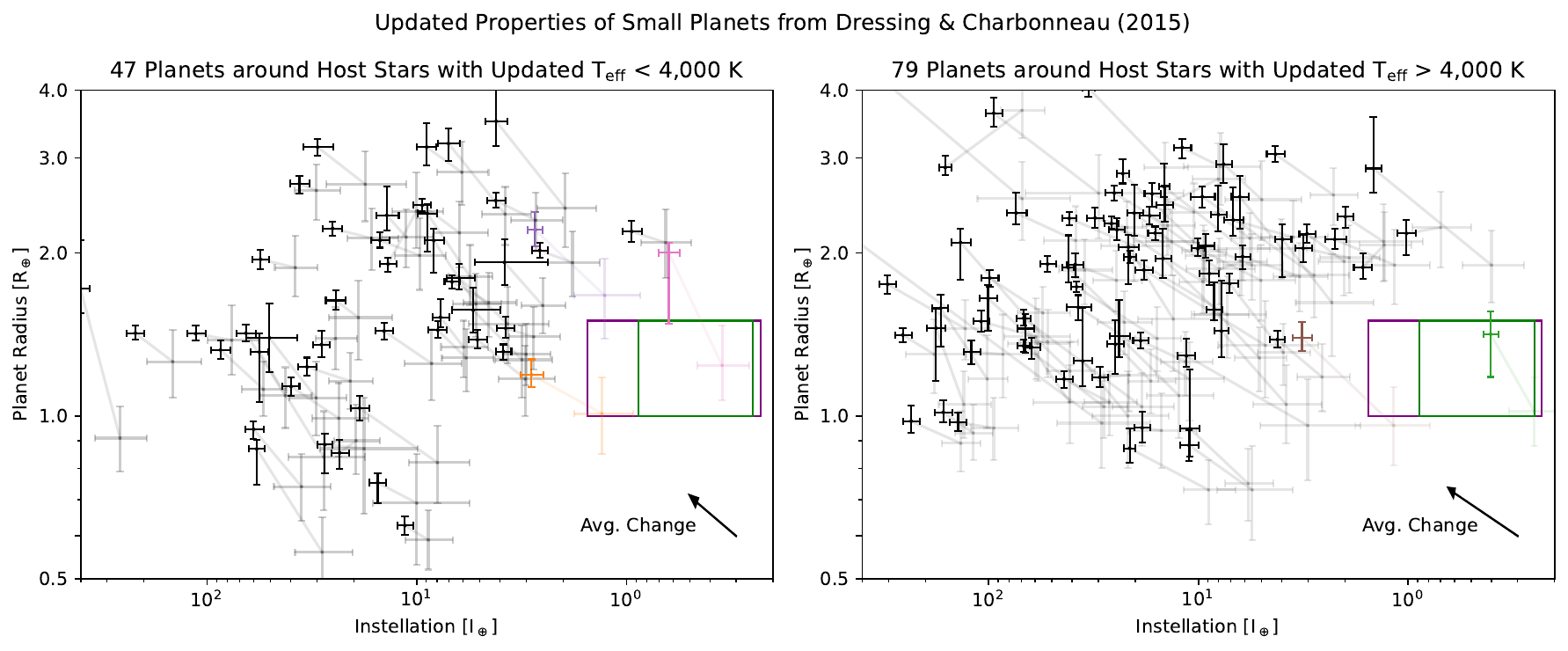}
    \end{minipage}
    \caption{\textbf{Top:} Original properties of small planets included in the \citet{DressingCharbonneau2015} study of \kepler{}'s M dwarfs ($T_\mathrm{eff} < 4000$\,K). Rectangles represent the optimistic (purple) and conservative (green) habitable zone boundaries used in \citet{DressingCharbonneau2015}. Translucent red points denote planets later labeled as false positives in \kepler{} \texttt{DR25} \citep{Thompson2018}, while opaque colored points represent ``Earth-sized" candidates considered relevant to the habitable zone. \textbf{Bottom:} current planet properties with updates from \gaia{} \texttt{DR2} \citep{GaiaDR2,Berger2020}, split between \textbf{left:} stars that are still below $4000$\,K and \textbf{right:} stars that are now warmer than $4000$\,K. Translucent points represent the original properties from \citet{DressingCharbonneau2015}, while opaque points denote the updated properties of those planets, with lines lines connecting the two for each individual planet. Black arrows denote the average ($\log_{10}$) change in planet properties, measured independently along either axis. The host star of one  of the six original habitable zone planets (K01681.04, blue) was not included in the \gaia{}-\kepler{} Stellar Properties Catalog \citep{Berger2020}, and thus does not appear in either of the updated parameter panels.}
    \label{fig:DC15}
\end{figure*}

Because the original \citetalias{DressingCharbonneau2015} sample was defined for cool stars with $T_\mathrm{eff} < 4000$\,K, it is important to note which stars did or did not meet this classification after receiving updated temperature estimates in \citet{Berger2020}. Thus, when plotting the \citetalias{DressingCharbonneau2015} planet sample with updated parameters in Figure~(\ref{fig:DC15}), we split the sample into those with host stars still considered cool, and those with host stars now considered too warm ($T_\mathrm{eff} > 4000$\,K) to be relevant. We found that two of the \citetalias{DressingCharbonneau2015} Earth-like planets are excluded from the cool star sample with this change (K00571.05 with $T_\mathrm{eff}=4023$\,K and K02650.01 with $T_\mathrm{eff}=4096$\,K). Of the still-cool stars, the aforementioned revisions of planetary parameters had caused the remaining three \citetalias{DressingCharbonneau2015} Earth-like planets (K00463.01, K02418.01, K03284.01) to no longer fall within the $[1,1.5]\,R_\oplus$, $[0.23,1.54]\,I_\oplus$ regime. 

The above changes left the original \citetalias{DressingCharbonneau2015} sample with no Earth-sized planets in the habitable zones of cool stars (lower left panel of Figure~\ref{fig:DC15}). We note that the \citetalias{DressingCharbonneau2015} study took place before the release of \kepler{}'s final catalog in \texttt{DR25}, such that it could not have made use of the full \kepler{} sample. More recently, \citet{Hsu2020} revisited this subject employing both \texttt{DR25} and their own revised stellar properties informed by \gaia{} and 2MASS (i.e., separate from the catalog of \citealp{Berger2020}). They found a similar paucity of relevant planets at habitable zone separations (see e.g., their Figure 3). More generally, they note that the uncertainties on \kepler{}'s M dwarf occurrence rate estimates are larger than previously believed: for example, for planets with $P < 50$\,days and $[1,2.5]$\,R$_\oplus$, \citetalias{DressingCharbonneau2015} found an occurrence rate of $\eta = 1.38_{-0.09}^{+0.11}$ while \citet{Hsu2020} found $\eta = 1.13_{-0.19}^{+0.20}$. 

While \citet{Hsu2020} provided an excellent evaluation of \kepler{}'s M dwarf occurrence rates, their work preceded an analysis on the importance of per-candidate reliability in \kepler{} demographic studies \citep{Bryson2020}. Reliability generally accounts for a candidate's odds of being a false alarm or false positive, and reliability scores (as defined in \citealp{Bryson2020-OG-Reliability}) are bounded between $0$ and $1$. In an occurrence calculation including reliability, a candidate's contribution will either remain at full weight (if perfectly reliable) or be downweighted (less reliable candidates contribute less), meaning that occurrence rates with reliability are typically lower than those without (see e.g., \citealp{Bryson2021, Bergsten2022}). Even in the scenario where a regime of interest is unpopulated (i.e., zero relevant candidates), accounting for reliability can help to ensure more robust extrapolations from populated regimes. As such, to make full use of available \kepler{} statistics, it is necessary to revisit the sample of M dwarfs incorporating \texttt{DR25}, \gaia{}-revised parameters, and candidate reliability in pursuit of updated occurrence rates.

\section{Current Sample} \label{sec:ModernSample}

\begin{figure}
    \centering
    \includegraphics[width=0.45\textwidth]{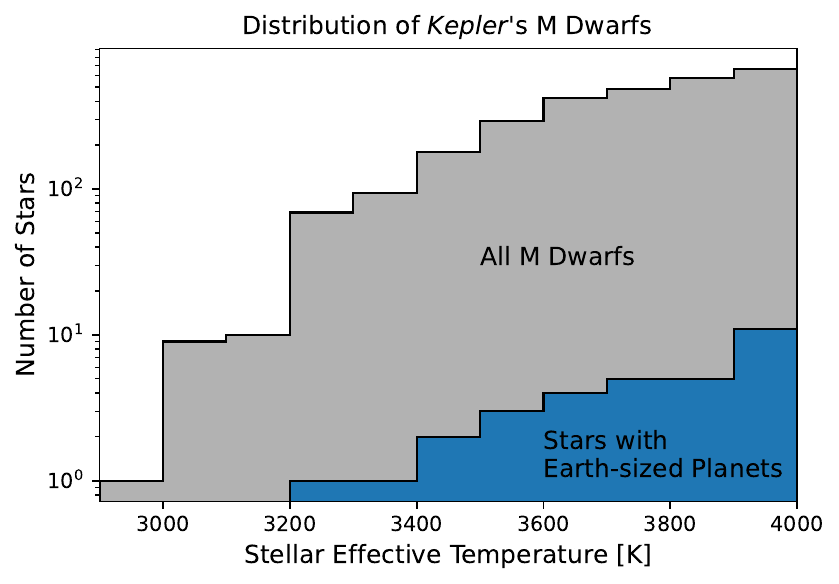}
    \caption{Distribution of stellar effective temperatures (from \citealp{Berger2020}) for our sample of \kepler{} M dwarf stars ($T_\mathrm{eff}<4000$\,K). The distribution of stars hosting Earth-sized planets is highlighted in blue, and is not significantly distinct from the overall distribution (a two-sample Kolmogorov-Smirnov provides a KS statistic of $0.16$ and a p-value of $0.38$).}
    \label{fig:stars}
\end{figure}

Having found the \citetalias{DressingCharbonneau2015} planet sample sufficiently changed from that of their original study, we elected to perform a new analysis using the current \kepler{} sample. Beginning from the \citet{Berger2020} catalog, we employed the same effective temperature cut ($T_\mathrm{eff} < 4000$\,K) used in \citetalias{DressingCharbonneau2015}. We also adopted the empirical selection criterion of \citet{Huber2016} where a star is considered a dwarf if:
\begin{equation}
    \log{g} > \frac{1}{4.671}\arctan{\left(\frac{T_\mathrm{eff}-6300}{-67.172}\right)}+3.876.
\end{equation}
This left a sample of $2807$ \kepler{} M dwarf stars; the distribution of stellar effective temperatures is shown in Figure~(\ref{fig:stars}). We then matched these stars with the \kepler{} \texttt{DR25} planet catalog to identify $60$ confirmed and $26$ candidate planets orbiting $62$ stars in this M dwarf sample, and calculated revised radii and instellations for these planets with the methodology described in the previous section. We also used the candidate reliability scores calculated in \citet{Bergsten2022}, which followed the approach of \citet{Bryson2020} to calculate reliability in terms of a candidate's false alarm reliability ($R_\mathrm{FA}$, \citealp{Bryson2020}) and false positive probability \citep{Morton2016}. All candidates in our sample had high reliability against being a false alarm ($R_\mathrm{FA}\gtrsim 95\%$), such that any low reliability scores were driven by a fairly high probability that a candidate was a false positive.

Previous works have indicated that the distribution of occurrence rates (and trends therein) evaluated in orbital period may differ from those evaluated in instellation for the same planet sample (see e.g., \citealp{Hsu2020, Petigura2022}). This may be especially prevalent when studying a small number of planets, where the dependence on individual stellar properties in converting between dimensions may introduce enough variation to alter the shape of the resulting occurrence distributions. To enable an assessment of how the chosen dimension might impact occurrence rate calculations, we defined two separate samples: one in orbital period, and one in instellation. For both samples, we adopted a radius range of $[0.5,1.5]\,R_\oplus$ to focus on Earth-sized planets. Adopting the same separation bounds as \citetalias{DressingCharbonneau2015} ($[0.5,200]$\,days; $[0.2,400]\,I_\oplus$) yielded a sample of $40$ planets around $32$ stars in orbital period and $39$ planets around $31$ stars in instellation \textit{when considering only median parameter values}. The difference in sample size is caused by one planet (K02542.01) with an instellation of $I=434_{-71}^{+64}\,I_\oplus$, though the errors are such that this planet may be included when accounting for input uncertainties.

\begin{figure*}[htb!]
    \centering
    \includegraphics[width=\textwidth]{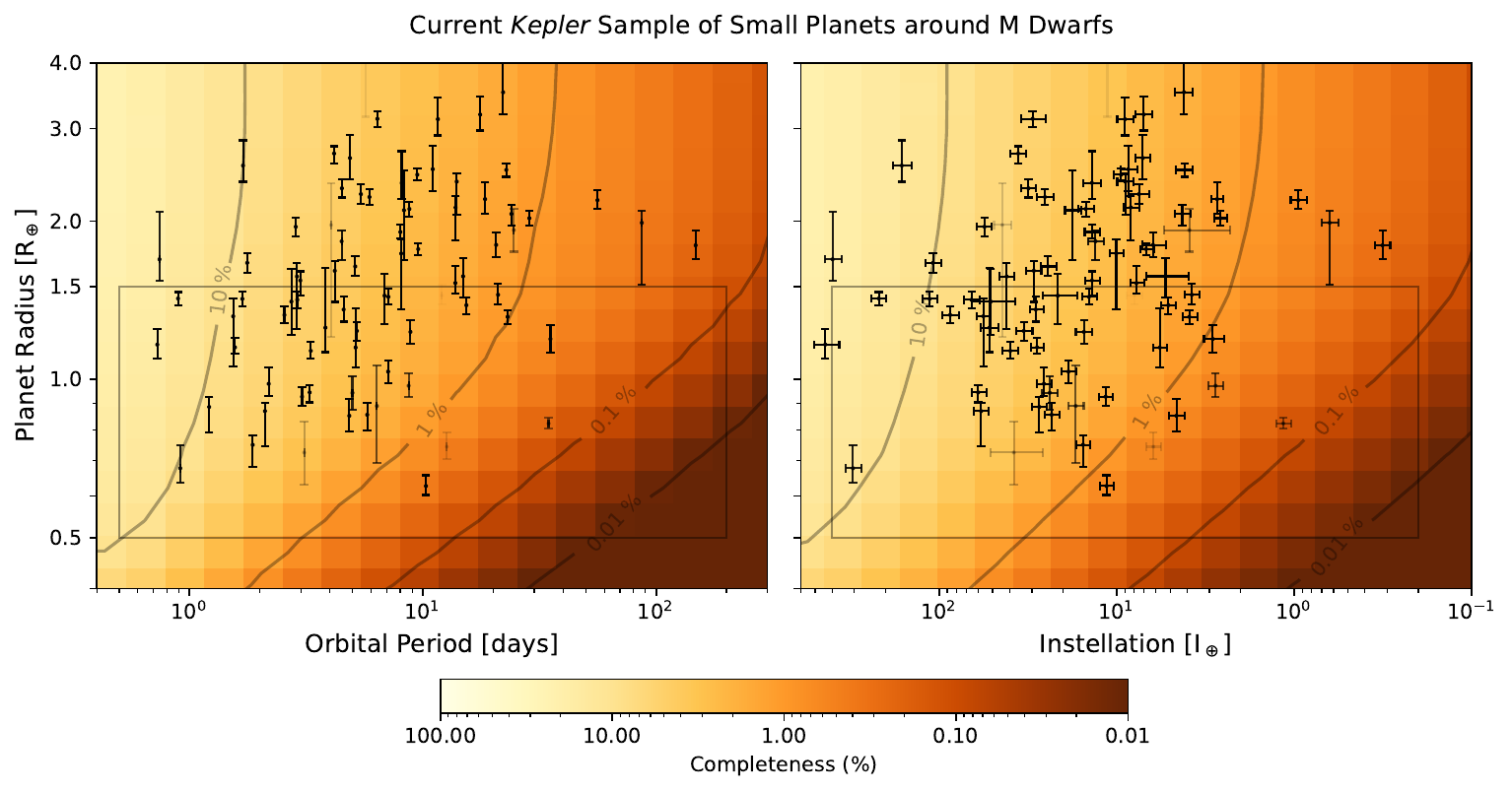}
    \caption{The current sample of \kepler{} small confirmed and candidate planets with \gaia{}-updated properties, plotted in \textbf{left:} orbital period and \textbf{right:} instellation. Error bars represent $1\sigma$ uncertainties, and opacity is determined by reliability (fully opaque corresponds to a reliability score of 1, \citealp{Bryson2020-OG-Reliability}). The average \kepler{} completeness maps for these M dwarfs are plotted in the background, with contours denoting relevant orders of magnitude ($0.01\%$ to $10\%$). Grey rectangles denote the parameter range(s) of interest for this study.}
    \label{fig:modern_samples}
\end{figure*}

In general, we note that sampling within parameter uncertainties can affect sample size and thus inferred occurrence rates. Relevant for this study, there are several planets whose median radii exceed $1.5\,R_\oplus$ but may fall into the Earth-sized classification within their uncertainties, including some with $P>50$\,days where we otherwise lack planets. As such, the samples shown in Figure~(\ref{fig:modern_samples}) include a larger number and range of planets than what is described above, as some may be included when we consider input uncertainties in Section~(\ref{sec:modeling}).


While the \kepler{} completeness contours used to calculate planet occurrence are defined in the planet radius/orbital period plane, we needed to convert these contours to the planet radius/instellation plane to calculate planet occurrence in instellation-space. Following the methodology of \citet{Bryson2021}, we took the per-star completeness maps\footnote{We created per-star completeness maps using a modified version of \texttt{epos} (\citealp{Mulders2018}, updated in \citealp{Bergsten2022}), which includes the per-star detection efficiency contours generated via \texttt{KeplerPORTs} \citep{BurkeCatanzarite2017}.} and applied a one-dimensional conversion, translating each orbital period point to a corresponding instellation flux value via Equations~(\ref{eqn:KeplersLaw}) and~(\ref{eqn:Flux}) using that star's median mass and luminosity from \citet{Berger2020}. We then interpolated each star's two-dimensional completeness map onto a uniform grid, and added the results for all stars to create the summed completeness grid necessary for Section~(\ref{sec:modeling}). Because the orbital period maps are all defined on the same grid, we may simply add them together as-is without re-interpolating onto a common grid. The average completeness maps (obtained by dividing the summed maps by the total number of stars) are shown in Figure~(\ref{fig:modern_samples}). 

\section{Occurrence Rate Methodology}\label{sec:Methods}

We adopted two approaches to calculating planet occurrence rates: direct evaluation via the inverse detection efficiency method, and indirect evaluation by way of a population forward model. The former is straightforward but can only constrain occurrence rates where there are exoplanet detections. Extrapolations from a forward model are thus needed to evaluate occurrence rates in poorly- or un-populated regimes, such as the longer-period (or lower-instellation) habitable zones that we were interested in studying. The inverse detection efficiency method is still a useful way of comparing model predictions in regimes where there is a significant number of exoplanet detections, and so we performed both evaluations where we could. 

All subsequent calculations were performed first in orbital period $P$ with the corresponding period sample, and then again in instellation $I$ (where $I$ was substituted for $P$ as needed) with the instellation sample.

\subsection{Inverse Detection Efficiency Method}

We employed the classical inverse detection efficiency method, modified to incorporate the reliability of planet candidates as in \citet{Bergsten2022}. For a survey with $N_*$ stars, the occurrence $\eta$ in a given bin of planetary parameters with $N_\mathrm{pl}$ planet candidates is given by:
\begin{equation}\label{eqn:IDEM}
    \eta_\mathrm{bin} = \frac{1}{N_*} \sum^{N_\mathrm{pl}}_{j} \frac{\mathrm{rel}_j}{\mathrm{comp}_j},
\end{equation}
with uncertainties given as $\sigma = \eta_\mathrm{bin}/\sqrt{N_\mathrm{pl}}$. Here, the $j^\mathrm{th}$ planet in a bin provides a weight proportional to its reliability $\mathrm{rel}_j$ and inversely proportional to the host star-specific completeness $\mathrm{comp}_j$ evaluated at that planet's parameters.

\subsection{Population Modeling}\label{sec:modeling}

Following the approach of \citet{Youdin2011} and \citet{Burke2015}, we adopted a population distribution function of the form:
\begin{equation}\label{eqn:pldf}
    \frac{\mathrm{d}^2 f}{\mathrm{d} P \mathrm{d} R} = F_\mathrm{0} C_\mathrm{n} g(P,R).
\end{equation}
The normalization factor $C_\mathrm{n}$ is defined such that the integral of Equation~(\ref{eqn:pldf}) over the entire sample domain (i.e., $[0.5,1.5]\,R_\oplus$ and $[0.5,200]$\,days or $[0.2,400]\,I_\oplus$) equals the scaling parameter $F_0$, which represents the average number of planets per star. The shape function $g(P,R)$ describes how the planet population behaves in orbital period and radius. Previous demographic studies have often made use of (broken) power laws in these dimensions to match observed occurrence rate distributions (see e.g., \citealp{Youdin2011, Howard2012,DongZhu2013,Burke2015}). We opted for a conventional broken power law in orbital period and a single power law in radius, defined by:
\begin{equation}\label{eqn:shape}
    g(P,R) = R^\alpha
    \begin{cases}
        (P / P_\mathrm{break})^{\beta_1} & \text{if $P < P_\mathrm{break}$} \\
        (P / P_\mathrm{break})^{\beta_2} & \text{if $P \geq P_\mathrm{break}$.}
    \end{cases}
\end{equation}
This contains four free parameters in addition to $F_0$: the exponents $\beta_1$ and $\beta_2$ control the slope of the occurrence distribution in orbital period space on either side of the break $P_\mathrm{break}$, and $\alpha$ determines the slope of the occurrence distribution in planet radius space.

The work of \citet{Bergsten2022} studied \kepler{}'s small planets around Sun-like (FGK) stars and found evidence of a period-dependent radius distribution that is not well-described by the simple power laws above. In their model, the small planet population shifts from being dominated by super-Earths at shorter periods to sub-Neptunes at longer periods, switching around some transition orbital period, and potentially signifying the effects of atmospheric mass loss processes. A large number of small planet detections was required to constrain the fairly complex functional form that described this behavior.

\citet{Bergsten2022} found the transition period to scale with stellar mass, and extrapolations of that trend would place the transition at very short orbital periods ($\sim5\pm1$\,days) for the median stellar mass of our sample ($0.50$\,M$_\odot$). This coincides with typical values of $P_\mathrm{break}$ in Equation~(\ref{eqn:shape}) for M dwarfs, such that the steep occurrence slope at shorter periods may explain the lack of detected planets needed to resolve the closest-in side of the transition. However, assuming a uniformly varying behavior across FGKM stars such that this extrapolation holds, this also means that the relative fractions of super-Earths and sub-Neptunes should be roughly flat with orbital period for the majority of our domain of interest. As such, the simple power laws of Equations~(\ref{eqn:pldf}) and~(\ref{eqn:shape}) may be suitable for modeling the current M dwarf sample especially given the presently limited number of detected planets; exploration of more complicated forms may be enabled by larger datasets in the future.

\subsubsection{Model Fitting}
The population model was fit to the planet sample by minimizing the corresponding Poisson likelihood function (Eqn.~9 in \citealp{Burke2015}) through a Markov-chain Monte Carlo process using the \texttt{emcee} \citep{ForemanMackey2012} package. Note that this likelihood function requires an evaluation of the sum of each star's completeness over an integral (or grid) of planet radii and orbital periods. We achieved this by evaluating each individual star's completeness on a uniform grid, and then taking the sum of all stars' completeness at each combination of planet parameters. We adopted uniform priors, and specifically fit the log of the power law break ($\log_{10}P_\mathrm{break}$) to ensure proper sampling over several orders of magnitude when repeating this process in instellation. 

We adopted three different methodologies to determine the input population and their parameters. In our first method (\texttt{M1}), we used the median values of each planetary parameter, thus considering only the confirmed and candidate planets with median values within the specified bounds. We used 64 walkers run for 20,000 steps and discard the first 1,000 for burn-in; justifications for optimization-related choices are included in Appendix~(\ref{app:mcmc}). While this method is conventional, it cannot account for the reliability of planet candidates nor the uncertainty in their input parameters (which themselves depend on the uncertainty of the stellar parameters used to calculate updated radii and instellations).

In our second method (\texttt{M2}), we followed the approach of \citet{Bryson2020-OG-Reliability} to implement reliability by performing separate inferences where the input planet population is drawn according to their reliability (e.g., a $50\%$ reliable candidate is included in $50\%$ of fits). We performed 100 inferences, each using 64 walkers run for 20,000 steps (discarding the first 1,000 for burn-in), and concatenated the posteriors to represent the global parameter distributions. 

In our third and most complete method (\texttt{M3}), we implemented both reliability and parameter uncertainties. For a given fit, we first took the entire planet sample around M dwarfs (not yet restricting to a specific parameter regime) and drew each candidate's parameters from within their (presumed independent) split-normal orbital period and radius distributions. If a candidate's drawn parameters were within the relevant regime, \textit{and} the candidate passed a separate draw according to its reliability (as in \texttt{M2}), then that candidate was included in the fit using the sampled parameters. We performed 400 of these inferences, each using 64 walkers run for 20,000 steps (discarding the first 1,000 for burn-in), and concatenated the posteriors.

While we sampled within the uncertainties of our planet parameters, a more inclusive approach may also sample within stellar effective temperature uncertainties to include stars (and their planets) near the $4000$\,K boundary. However, as noted in \cite{Bryson2021}, because the number of stars would vary between inferences, the summed completeness components would also need re-evaluation for each inference, which is computationally expensive and beyond the scope of this work.

\section{Results}\label{sec:Results}

\begin{figure*}[htb!]
    \centering
    \includegraphics[width=\textwidth]{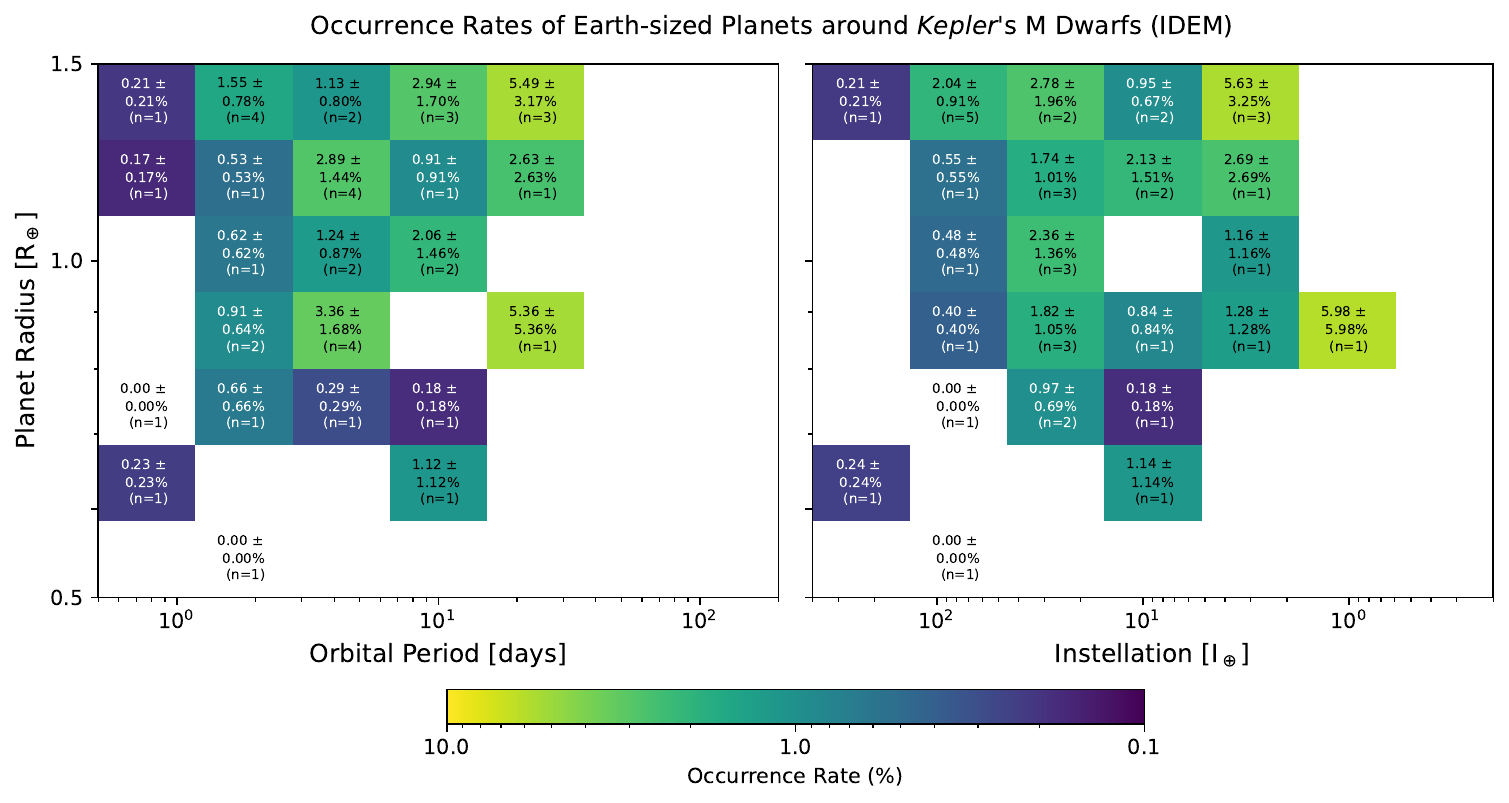}
    \caption{Occurrence rates of Earth-sized planets around M dwarfs distributed in \textbf{left:} orbital period and \textbf{right:} instellation, calculated via the reliability-weighted inverse detection efficiency method (Eqn.~\ref{eqn:IDEM}). White bins with a printed occurrence of $\eta=0\%$ indicate bins that are technically populated, but by a candidate(s) with a reliability of 0.}
    \label{fig:IDEM}
\end{figure*}

\begin{figure*}[htb!]
    \centering
    \includegraphics[width=\textwidth]{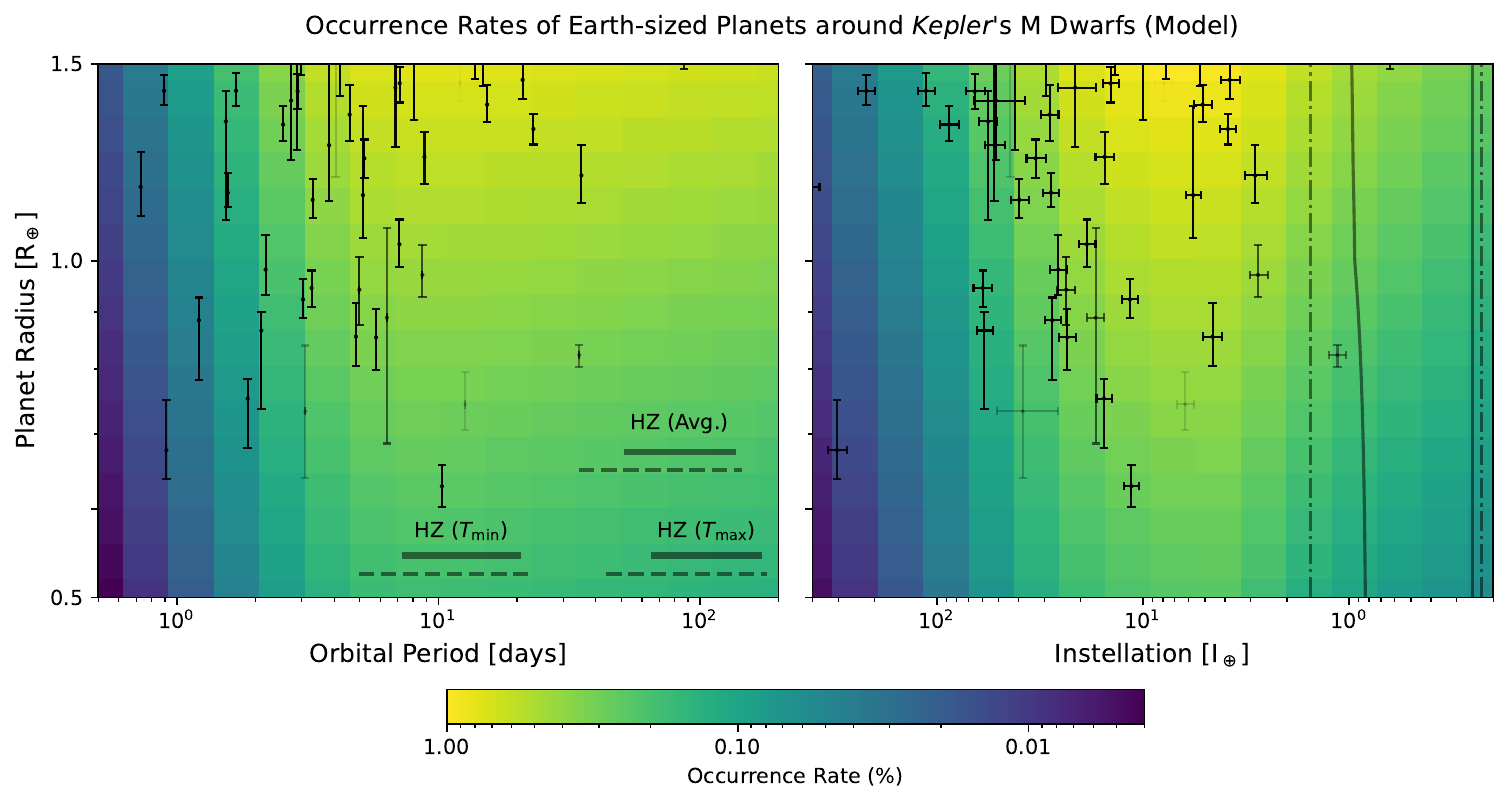}
    \caption{Modeled occurrence rates of Earth-sized planets around M dwarfs distributed in \textbf{left:} orbital period and \textbf{right:} instellation, calculated by evaluating Equation~(\ref{eqn:pldf}) with the \texttt{emcee}-output parameter vectors from method \texttt{M3} (plotted values represent median occurrence rates). Planet candidates from Figure~(\ref{fig:modern_samples}) are reproduced here as points with error bars denoting their $1\sigma$ parameter uncertainties. Dashed gray lines represent the optimistic habitable zone, and solid gray lines represent the conservative habitable zone \citep{Kopparapu2013, Kopparapu2014}. The three cases in orbital period represent the habitable zone bounds for the coldest ($T_\mathrm{min}\approx2979$\,K) and warmest ($T_\mathrm{max}\approx4000$\,K) stars in our sample, along with the average of the bounds for all stars in our sample.}
    \label{fig:model_grids}
\end{figure*}

\begin{figure*}[htb!]
    \centering
    \includegraphics[width=\textwidth]{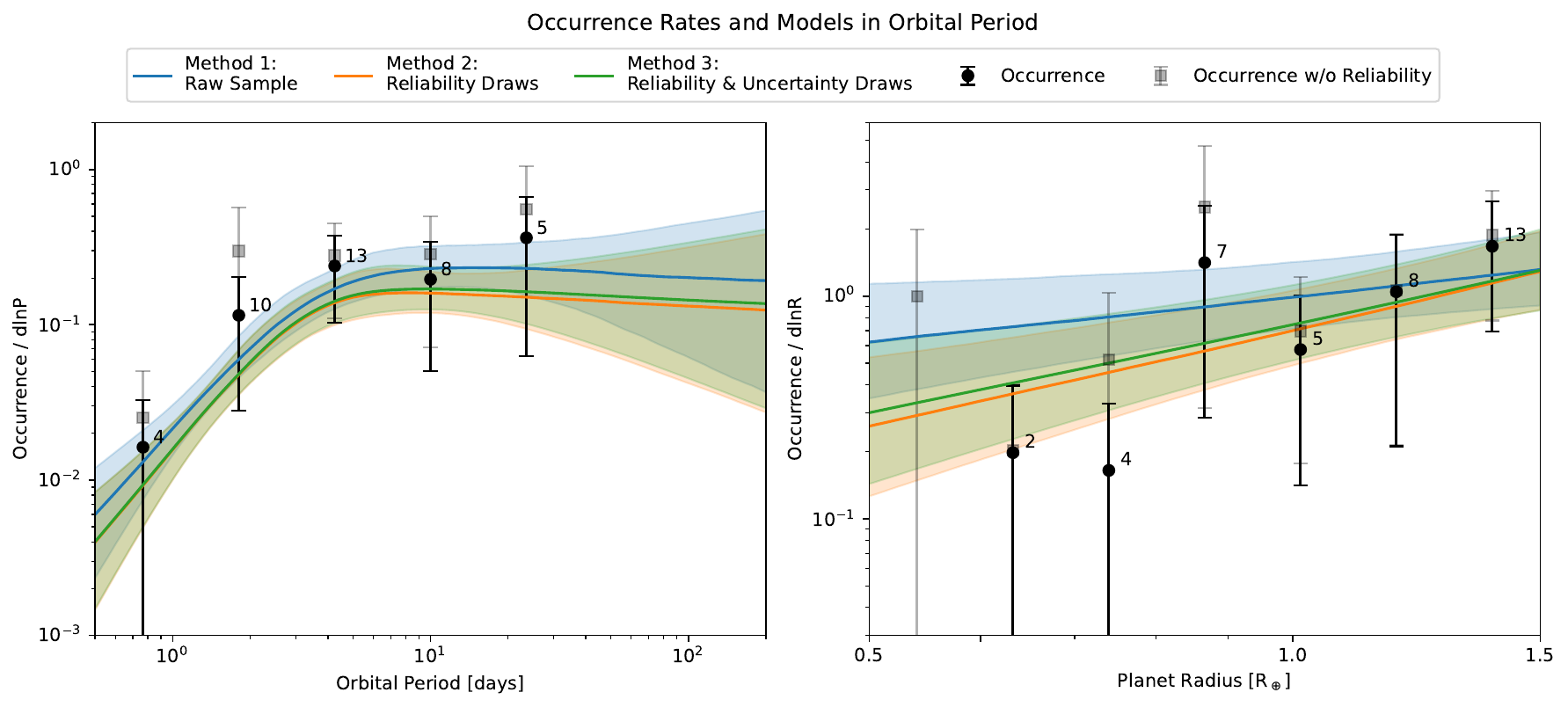}
    \caption{Best-fit population models and observed occurrence rates marginalized to show distributions in \textbf{left:} orbital period and \textbf{right:} planet radius. Individual points represent occurrence rates calculated via the inverse detection efficiency method (Eqn.~\ref{eqn:IDEM}), with and without accounting for reliability (black and gray points, respectively); numbers represent the number of candidates within each bin. For each method, the colored curve represents the median of the distribution of occurrence rates found by evaluating the population model with the distribution of parameter vectors; lighter filled regions denote the $1\sigma$ uncertainty envelopes. Colors represent the results of individual fitting methodologies: \texttt{M1} (blue) considers only the median parameter values of planets within $[0.5,200]$\,days, $[0.5,1.5]\,R_\oplus$ and treats all planets as fully reliable; \textit{M2} (orange) is similar to \texttt{M1} but repeatedly draws planets based on their reliability and concatenates the posteriors; \textit{M3} (green) is similar to \textit{M2} but also draws each planet's parameters from within their uncertainties, allowing for the inclusion of planets with median values outside the designated range.}
    \label{fig:model_P}
\end{figure*}

\begin{figure*}[htb!]
    \centering
    \includegraphics[width=\textwidth]{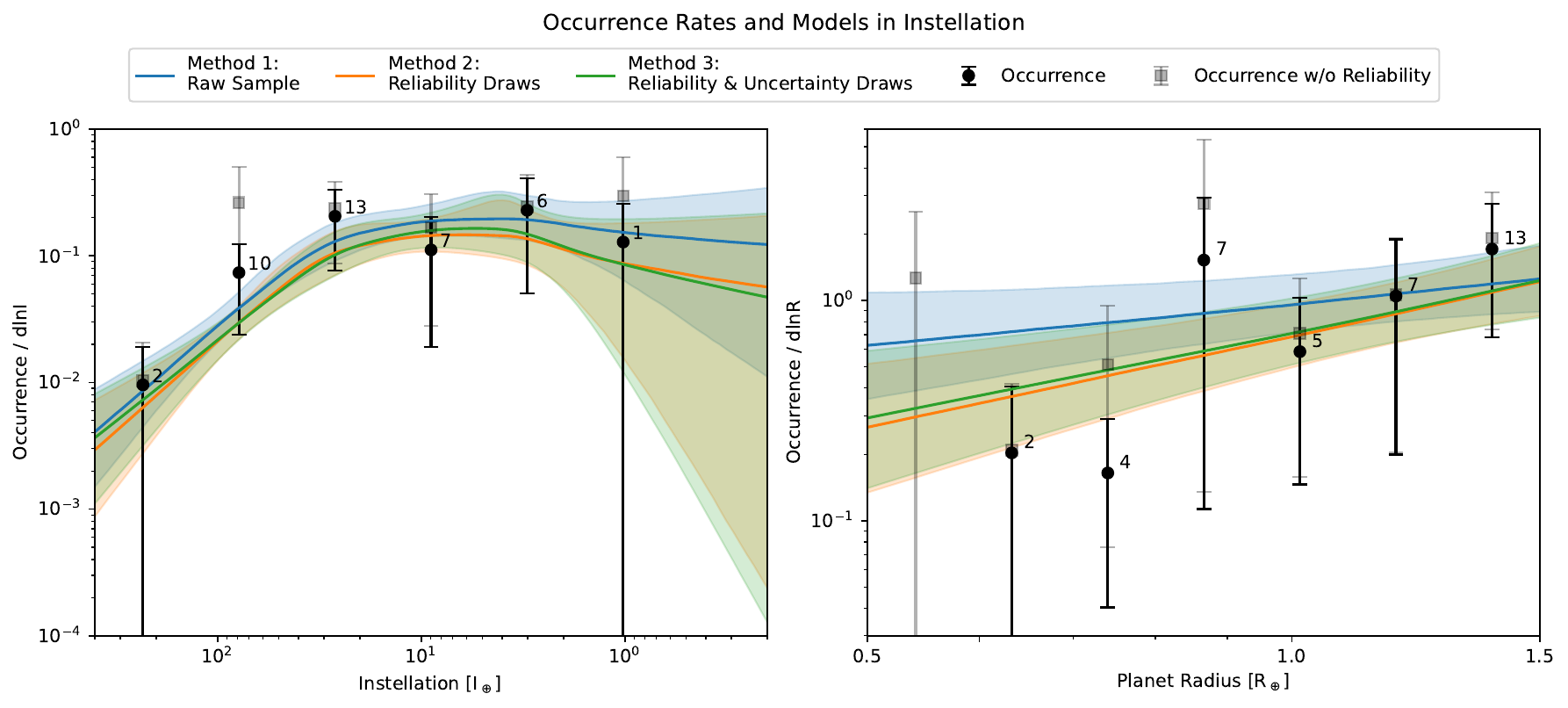}
    \caption{Similar to Figure~(\ref{fig:model_P}) but with evaluations in instellation (rather than orbital period) for the range of $[0.2,400]\,I_\oplus$, $[0.5,1.5]\,R_\oplus$.}
    \label{fig:model_F}
\end{figure*}

\begin{table*}[]
    \begin{tabular}{c|c|ccccc}
        Dimension & Method & $F_0$ & $\alpha$ & $P_\mathrm{break}$ & $\beta_1$ & $\beta_2$ \\ \hline
        \multirow{3}{*}{Orbital Period} & \texttt{M1} & ${1.04}_{-0.30}^{+0.47}$ & ${-0.32}_{-0.58}^{+0.61}$ & ${4.10}_{-2.03}^{+4.55}$ & ${0.76}_{-0.55}^{+1.00}$ & ${-1.06}_{-0.69}^{+0.37}$\\
         & \texttt{M2} & ${0.72}_{-0.22}^{+0.37}$ & ${0.45}_{-0.69}^{+0.74}$ & ${3.63}_{-1.57}^{+2.84}$ & ${0.90}_{-0.57}^{+1.03}$ & ${-1.08}_{-0.55}^{+0.37}$\\
         & \texttt{M3} & $\mathbf{{0.77}_{-0.24}^{+0.39}}$ & $\mathbf{{0.33}_{-0.73}^{+0.77}}$ & $\mathbf{{3.64}_{-1.57}^{+3.10}}$ & $\mathbf{{0.91}_{-0.57}^{+1.02}}$ & $\mathbf{{-1.07}_{-0.58}^{+0.37}}$\\
        [0.5ex]\hline\hline
         & & & & $I_\mathrm{break}$ & &\\ \hline
        \multirow{3}{*}{Instellation} & \texttt{M1} & ${1.01}_{-0.26}^{+0.39}$ & ${-0.37}_{-0.60}^{+0.62}$ & ${23.17}_{-18.60}^{+22.94}$ & ${-0.89}_{-0.29}^{+1.05}$ & ${-2.38}_{-0.63}^{+0.48}$\\
         & \texttt{M2} & ${0.70}_{-0.19}^{+0.29}$ & ${0.38}_{-0.69}^{+0.74}$ & ${23.04}_{-19.31}^{+21.87}$ & ${-0.76}_{-0.35}^{+2.40}$ & ${-2.43}_{-0.73}^{+0.56}$\\
         & \texttt{M3} & $\mathbf{{0.73}_{-0.20}^{+0.33}}$ & $\mathbf{{0.30}_{-0.75}^{+0.79}}$ & $\mathbf{{18.20}_{-14.61}^{+18.13}}$ & $\mathbf{{-0.67}_{-0.43}^{+2.58}}$ & $\mathbf{{-2.27}_{-0.69}^{+0.44}}$\\
         \hline
    \end{tabular}
    \caption{Median and $1\sigma$ (\nth{16} and \nth{84} percentile) parameter values for the optimized population models in orbital period and instellation. Rows in bold are from our most complete \texttt{M3} method which includes both reliability and parameter uncertainties. While we fit the log of the break parameters (i.e., $\log_{10} P_\mathrm{break}$ and $\log_{10} I_\mathrm{break}$), we include the corresponding linear values here for readability. Note that because Equation~(\ref{eqn:pldf}) is defined in terms of $\mathrm{d} P$ and $\mathrm{d} R$, a value of $-1$ in the power law exponents ($\alpha,\beta_1,\beta_2$) corresponds to a flat line in natural log occurrence (i.e., $\frac{\mathrm{d}P}{\mathrm{d}\ln{P}} = P$). Additionally, because orbital period increases with distance from a star while instellation decreases, the far-out regime (beyond the break) is governed by $\beta_2$ in orbital period but by $\beta_1$ in instellation.}
    \label{tab:param}
\end{table*}

The occurrence rate grids from the reliability-weighted inverse detection efficiency method are presented in Figure~(\ref{fig:IDEM}) for both orbital period and instellation. These occurrence rates were then marginalized along either axis to provide the one-dimensional occurrence rates seen in Figure~(\ref{fig:model_P}) for orbital period and Figure~(\ref{fig:model_F}) for instellation. To determine the distribution of occurrence rates predicted by our population model(s), we evaluated Equation~(\ref{eqn:pldf}) with the \texttt{emcee}-output parameter vectors to produce occurrence grids like those in Figure~(\ref{fig:model_grids}), and further integrated along either axis to present the modeled marginalized distributions in Figures~(\ref{fig:model_P}) and~(\ref{fig:model_F}). The best-fit parameter values are presented in Table~(\ref{tab:param}), where we treated the \nth{16} and \nth{84} percentiles as $1\sigma$ bounds about the median (\nth{50} percentile) value.

We found values from the inverse detection efficiency method to fall within model predictions for results in both orbital period and instellation (left panels of Figures~\ref{fig:model_P} and~\ref{fig:model_F}, respectively). For the \texttt{M2} and \texttt{M3} methods which incorporate candidate reliability, we found that the models from both methods exhibit $1\sigma$ agreement with the inverse detection efficiency method in all parameter bins where there are planet candidates (such that Equation~\ref{eqn:IDEM} may be evaluated). For the \texttt{M1} case which does not consider reliability, we re-evaluated Equation~(\ref{eqn:IDEM}) while neglecting the $\mathrm{rel}_j$ term (i.e., the classical definition of the inverse detection efficiency method) and found that these values agree with the \texttt{M1} model predictions within $1\sigma$ for all populated bins.

The radius distribution of occurrence values (right panels of Figures~\ref{fig:model_P} and~\ref{fig:model_F}) between the models and the inverse detection efficiency method are typically consistent within $1\sigma$. However, we note that smaller radius bins require $\sim 1.1\sigma$ to achieve consistency, and that there is appreciable variation between methods (compared to what is seen with the period or instellation distributions). We attribute these caveats to our small sample size and correspondingly sparse coverage when splitting the sample into bins of planet radius to evaluate with the inverse detection efficiency method. We further note that in the smallest planet radius bin ($\lesssim 0.6\,R_\oplus$), there is only one planet and it has a reliability score of zero, meaning that Equation~(\ref{eqn:IDEM}) returns a zero occurrence rate (as shown in Figure~\ref{fig:IDEM}).

Regarding differences between each of the model fitting methodologies, we found that (for a given dimension) model parameters are consistent at the $1\sigma$ level across all methods. There is a slight (but expected) global decrease in predicted occurrence between \texttt{M1} and \texttt{M2} characteristic of reliability incorporation, where the observed number of planets decreases due to some candidates being considered false positives or false alarms in a given inference. As in \citet{Bryson2020-OG-Reliability}, we found that input uncertainties have little effect when accounting for reliability, and thus the parameters between models fit via \texttt{M2} and \texttt{M3} are very similar. As there is no appreciable difference, we favor results from the more-complete \texttt{M3} approach incorporating both reliability and input uncertainties.

\begin{figure}
    \centering
    \includegraphics[width=0.45\textwidth]{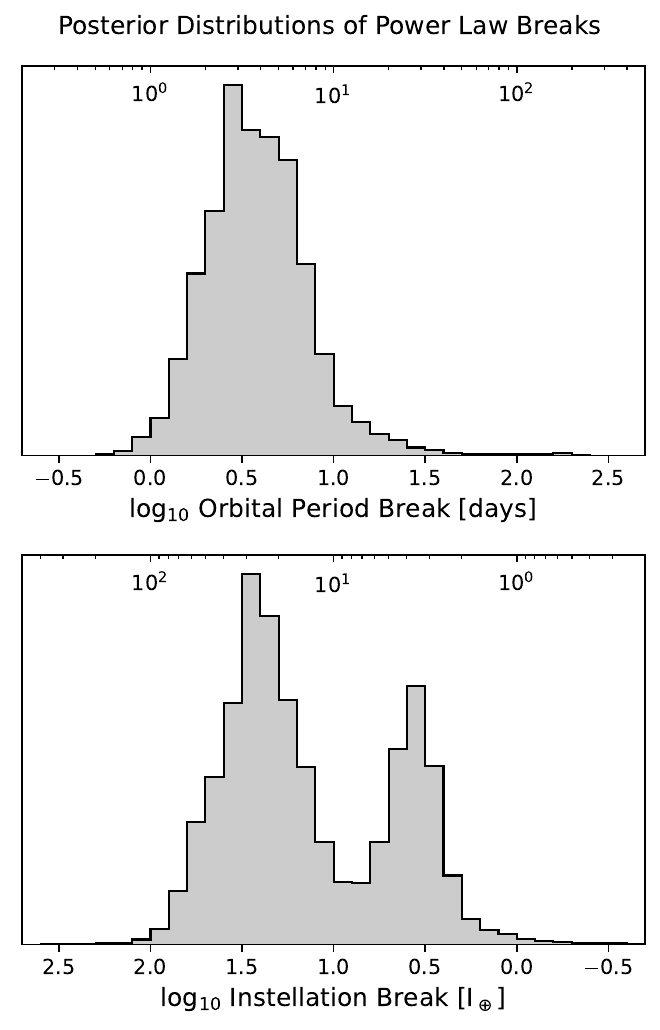}
    \caption{Posterior distributions for the $\mathrm{log}_{10}$ power law breaks in \textbf{top:} orbital period and \textbf{bottom:} instellation from the \texttt{M3} models. Tick marks on the top of each panel represent the corresponding non-logarithmic values. There is a bimodality in the instellation break posteriors, meaning our model parameters are not well-constrained in this dimension; this feature is discussed further in Appendix~(\ref{app:bimodality}).}
    \label{fig:breaks}
\end{figure}

We note that the uncertainty contours on the population model results for the occurrence distribution in instellation (left panel of Figure~\ref{fig:model_F}) show a secondary bump around $\sim4\,I_\oplus$ separate from the power law's main break around $\sim25\,I_\oplus$. In Figure~(\ref{fig:breaks}) we highlight that the posterior distribution for the break parameter in instellation (modeled as $\log_{10} I_\mathrm{break}$) appears to follow a bimodal distribution, and is thus not well-represented by summary statistics that assume a unimodal (split normal) distribution. This is not true of the models in orbital period as the $\log_{10} P_\mathrm{break}$ distribution appears unimodal. We explore the possibility of the instellation bimodality representing physically distinct populations in Appendix~(\ref{app:bimodality}), but ultimately conclude the feature is likely an artifact of our small sample size and sparse distribution. Nevertheless, having a model defined in instellation greatly simplifies evaluations of the habitable zone (Section~\ref{sec:HZ}) and enables comparisons across spectral types (Section~\ref{sec:FGK}). As such, we consider and compare occurrence estimates from both our period and instellation models where applicable throughout the remainder of this work.

\subsection{Habitable Zone Occurrence Rates}\label{sec:HZ}

\begin{figure*}[htb!]
    \centering
    \includegraphics[width=\textwidth]{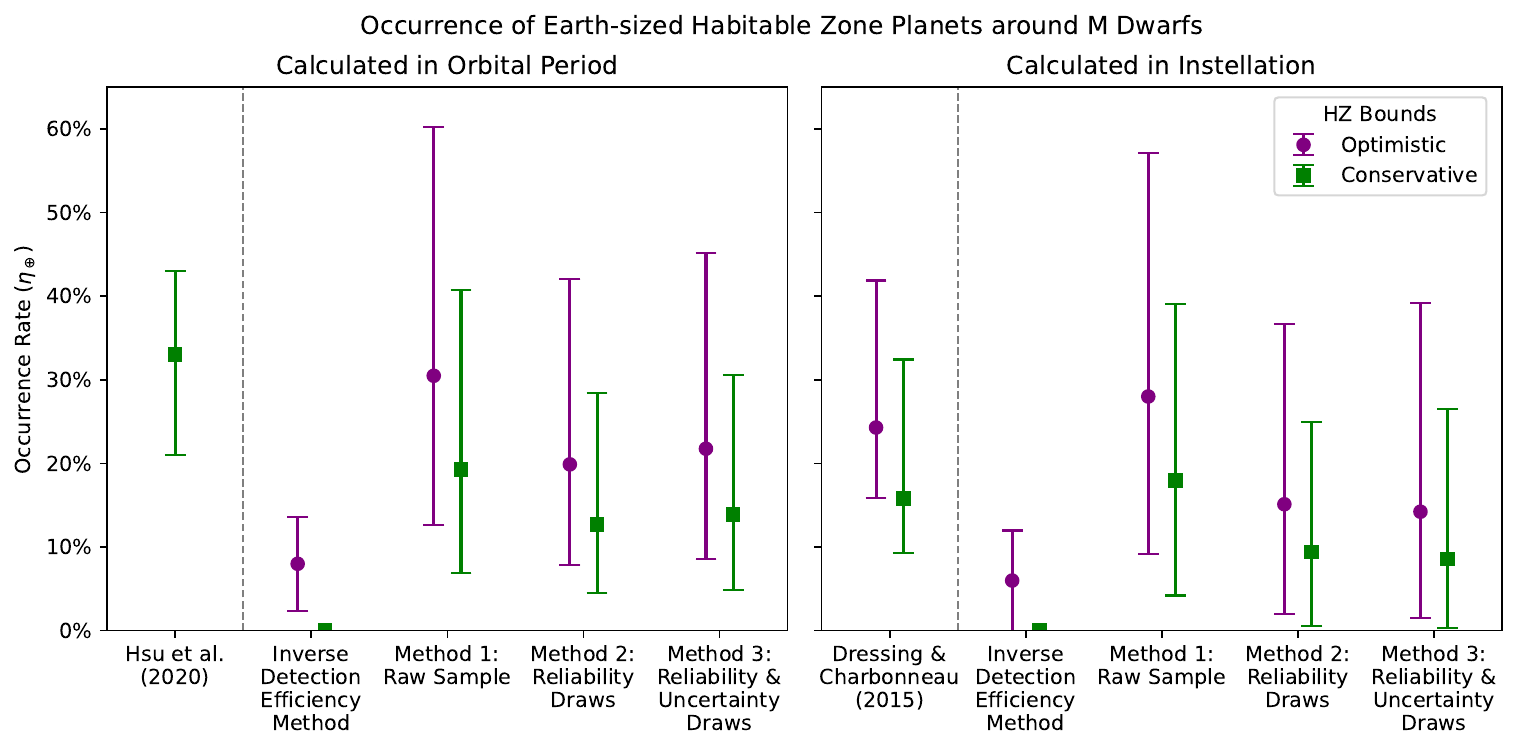}
    \caption{The occurrence rate of Earth-sized habitable zone planets around M dwarfs calculated in this work in \textbf{left:} orbital period and \textbf{right:} instellation. In either dimension, we include a comparison to previous results: the modeled results of \citet{Hsu2020} calculated in orbital period, or the results of \citet{DressingCharbonneau2015} calculated in instellation with the inverse detection efficiency method. Purple and green points denote calculations for optimistic and conservative habitable zone bounds, respectively. For our plotted values in orbital period, we adopted the average habitable zone bounds for the stars in our sample; different stellar effective temperature cases are considered in Table~(\ref{tab:etaEarth}).
    }
    \label{fig:etaEarth}
\end{figure*}

We used the habitable zone models of \citet{Kopparapu2013} to define the optimistic (recent Venus out to early Mars) and conservative (runaway greenhouse out to maximum greenhouse) boundaries. In Appendix~(\ref{app:mass-dependent-HZ}), we describe our considerations of a planet-mass-dependent modulation to the conservative inner (i.e., runaway greenhouse) edge. The \citet{Kopparapu2013} habitable zone bounds are near-constant in flux across the stellar effective temperature range of M dwarfs: the conservative inner and outer edges vary by $\sim3\%$ and $\sim12\%$ respectively between the warmest and coolest stars in our sample; the optimistic bounds similarly vary by $\sim4\%$ and $\sim14\%$. Due to this consistency, we are unlikely to benefit from a three-dimensional model that includes a stellar effective temperature dependence alongside instellation and planet radius, as was used in \citet{Bryson2021} for Sun-like stars where the habitable zone varies more strongly with temperature.

However, the habitable zone bounds have a considerable range when converted to the corresponding orbital periods: the conservative innermost (outermost) edge falls at roughly $8$ ($20$) days for our coolest star ($T_\mathrm{min}\approx2979$\,K), but at $67$ ($168$) days for our warmest star ($T_\mathrm{max}\approx4000$\,K). Because our stellar sample's effective temperature distribution is clustered towards hotter stars (Figure~\ref{fig:stars}), the average conservative habitable zone boundaries of $52$ ($133$) days are closer to the latter. We considered these three cases in our habitable zone calculations (for both conservative and optimistic bounds) in orbital period. While a three-dimensional model in orbital period, planet radius, and stellar effective temperature would have been useful here, fitting such a model would have required a greater number and coverage of systems than what is available with the current \kepler{} sample (see Appendix~\ref{app:TeffSplit}).


To calculate each model's Earth-sized habitable zone planet occurrence rate ($\eta_\oplus$), we evaluated an integral of Equation~(\ref{eqn:pldf}) over the respective bounds (conservative and optimistic) for the three temperature cases in orbital period, and for the average case in instellation. We considered the range of \texttt{emcee}-output parameter vectors, and report the median values with $1\sigma$ uncertainties in Tables~(\ref{tab:etaEarth}). These values are also plotted in Figure~(\ref{fig:etaEarth}) alongside previous estimates from \citetalias{DressingCharbonneau2015} who used the inverse detection efficiency method (without the $\mathrm{rel}_j$ term), and from \citet{Hsu2020} who used a forward model without per-candidate reliability. We stress that, due to the paucity of observed habitable zone planets, our results are predominantly \textit{model-driven extrapolations}; a purely observational result derived from the reliability-weighted inverse detection efficiency method (Equation~\ref{eqn:IDEM}) is also included for comparison.


Taking the \texttt{M3} orbital period model results evaluated over $[0.5, 1.5]$\,R$_\oplus$ with the average-temperature habitable zone bounds, we found $\eta_\oplus={13.89}_{-9.02}^{+16.70}\%$ for the conservative case (and ${21.75}_{-13.23}^{+23.44}\%$ for the optimistic). For the \texttt{M3} instellation model, we found $\eta_\oplus = {8.58}_{-8.22}^{+17.94}\%$
(${14.22}_{-12.71}^{+24.96}\%$). These occurrence rates are consistent at the $1\sigma$ level with each other, the estimates of $15.82^{+16.60}_{-6.54}\%$ ($24.28^{+17.58}_{-8.39}\%$) for $[1.0,1.5]\,R_\oplus$ planets from \citetalias{DressingCharbonneau2015}, and the larger prediction of $33^{+10}_{-12}\%$ for $[0.75,1.5]\,R_\oplus$ planets in the conservative habitable zone from \citet{Hsu2020}.\footnote{Our model estimates are also consistent with literature estimates when using their unique radius ranges, included here for posterity. Evaluating our \texttt{M3} models over $[1.0,1.5]\,R_\oplus$ gives $7.41^{+8.59}_{-4.78}\%$ ($11.60^{+12.00}_{-7.01}\%$) in orbital period and $7.54^{+12.74}_{-6.73}\%$ ($8.58^{+17.94}_{-8.22}\%$) in instellation, all consistent with the estimates from \citetalias{DressingCharbonneau2015}. Likewise, using $[0.75,1.5]\,R_\oplus$ gives $10.78^{+12.56}_{-6.97}\%$ ($16.88^{+17.58}_{-10.21}\%$) in orbital period and $8.58^{+17.94}_{-8.22}\%$ ($11.00^{+18.70}_{-9.83}\%$) in instellation, consistent with \citet{Hsu2020}.} Our results are further consistent with the upper limit of $\eta_\oplus < 23\%$ set by \citet{Pinamonti2022} using radial velocity data.

We note that our uncertainties tend to be slightly larger than previous works: in the conservative case, the $1\sigma$ uncertainties from \citetalias{DressingCharbonneau2015} span $\sim 9-32\%$ and the estimate from \citet{Hsu2020} spans $\sim21-43\%$, while our \texttt{M3} orbital period estimate spans $\sim5-31\%$ and the instellation estimate spans $\sim0-27\%$. We attribute this increased uncertainty to the smaller number of habitable zone planet candidates and our subsequent need to rely on extrapolated models. The latter requires assumptions of a functional form and that the population continues to behave in some uniformly varying way across the extrapolated regime, which adds an inherent uncertainty not reflected in the statistical error bars of Figure~(\ref{fig:etaEarth}). As such, despite comparable results to previous works, our new exploration of the \gaia{}-informed \kepler{} sample leads us to conclude that \textbf{the occurrence of Earth-sized habitable zone planets around M dwarfs is not as observationally constrained as previously thought}. 

\begin{table*}[!ht]
    \begin{tabular}{c|ccc|ccc}
    \multicolumn{7}{c}{Occurrence of Earth-sized Habitable Zone Planets ($\eta_\oplus$, $\%$)}\\\hline\hline
        \multicolumn{7}{c}{Orbital Period}\\\hline
         & \multicolumn{3}{c|}{Optimistic} & \multicolumn{3}{c}{Conservative}\\\hline
         & $T_\mathrm{min}$ & Average & $T_\mathrm{max}$ & $T_\mathrm{min}$ & Average & $T_\mathrm{max}$ \\ 
        Bounds [days] & ${5}-{22}$ & ${34}-{144}$ & ${44}-{181}$ & ${8}-{20}$ & ${52}-{133}$ & ${67}-{168}$ \\\hline
        Observed Planets & 14 & 2 & 0 & 6 & 0 & 0 \\ \hline
        IDEM & ${13.83} \pm {3.70}$ & ${8.00} \pm {5.65}$ & -- & ${4.92} \pm {2.01}$ & -- & -- \\
        \texttt{M1} & ${35.11}_{-8.47}^{+11.59}$ & ${30.47}_{-17.86}^{+29.77}$ & ${29.36}_{-19.05}^{+34.02}$ & ${23.47}_{-5.75}^{+8.90}$ & ${19.28}_{-12.41}^{+21.47}$ & ${18.70}_{-13.05}^{+24.05}$ \\
        \texttt{M2} & ${24.41}_{-6.13}^{+8.11}$ & ${19.88}_{-11.99}^{+22.13}$ & ${19.27}_{-12.55}^{+25.14}$ & ${15.97}_{-4.16}^{+5.66}$ & ${12.70}_{-8.19}^{+15.76}$ & ${12.25}_{-8.42}^{+17.70}$ \\
        \texttt{M3} & ${25.91}_{-6.63}^{+8.93}$ & ${21.75}_{-13.23}^{+23.44}$ & ${21.05}_{-13.79}^{+26.59}$ & ${17.11}_{-4.61}^{+6.36}$ & ${13.89}_{-9.02}^{+16.70}$ & ${13.38}_{-9.26}^{+18.79}$ \\\hline\hline
        \multicolumn{7}{c}{Instellation}\\\hline
         & \multicolumn{3}{c|}{Optimistic} & \multicolumn{3}{c}{Conservative}\\\hline
        Bounds [$I_\oplus$] & & ${0.23}-{1.54}$ & & & ${0.25}-{0.88}$\textsuperscript{a} & \\ \hline
        IDEM & & ${6.01} \pm {6.01}$ & & & -- & \\ 
        \texttt{M1} & & ${27.99}_{-18.87}^{+29.14}$ & & & ${17.92}_{-13.72}^{+21.10}$ & \\
        \texttt{M2} & & ${15.11}_{-13.15}^{+21.59}$ & & & ${9.42}_{-8.89}^{+15.54}$ & \\
        \texttt{M3} & & $\mathbf{{14.22}_{-12.71}^{+24.96}}$ & & & $\mathbf{{8.58}_{-8.22}^{+17.94}}$ & \\\hline
    \end{tabular}
    \caption{Earth-sized habitable zone occurrence ($\eta_\oplus$, in $\%$) estimates for M dwarfs calculated in orbital period and instellation. Because the habitable zone in orbital period varies greatly with spectral type, we provide three cases for either the optimistic or conservative bounds: the bounds for the coolest ($T_\mathrm{min}\approx2979$\,K) or warmest ($T_\mathrm{max}\approx4000$\,K) stars in our sample, and the average bounds across all stars in our sample. The habitable zone in instellation is roughly constant with spectral type, so we provide only one case evaluated for the average stellar effective temperature of our sample. \small\textsuperscript{a}Our incorporation of a planet radius dependence to the runaway greenhouse edge causes this value to vary from $0.83\,I_\oplus$ at $R=0.5\,R_\oplus$ to $0.97\,I_\oplus$ at $R=1.5\,R_\oplus$.}
    \label{tab:etaEarth}
\end{table*}


\section{Comparisons across Spectral Types}\label{sec:FGK}

Previous studies have explored \kepler{}'s small planets for dependencies on stellar mass across the range of FGKM stars. \citet{Mulders2015} studied the distribution of small ($1-4$\,R$_\oplus$) planet occurrence rates with semimajor axis for different spectral types using the planet catalog of \citet{Burke2014}. They found that planet occurrence increased towards later spectral types, rising by a factor of $\sim2$ from G to M stars and by a factor of $\sim3$ from F to M stars. \citet{Mulders2015b} used the \kepler{} Q1-Q16 planet candidate sample \citep{Mullally2015} to find a factor of $\sim3.5$ increase from FGK to M stars for small ($1-2.8$\,R$_\oplus$) planets while also noting a factor of $\sim2$ decrease for larger ($>2.8$\,R$_\oplus$) planets. These estimates were based on the close-in sample ($P<50$\,days), but have been used to predict $\eta_\oplus$ around M dwarfs by upscaling estimates from FGK stars (see e.g., \citealp{KHU2023}), despite (a) the habitable zone falling at much longer orbital periods for the latter and (b) the inclusion of planets larger than $1.5$\,R$_\oplus$.

However, recent work suggests the factor of $\sim3.5$ increase around M dwarfs may not be applicable to habitable zone occurrence rates. Employing the final \texttt{DR25} sample of \kepler{} candidates along with \gaia{}-revised properties, \citet{Hsu2020} found that M dwarfs have higher occurrence rates than FGK stars \citep{Hsu2019} at the same orbital periods, but that the two groups have comparable occurrence rates when evaluated at similar instellations. For example, they reported an M/FGK ratio of $3.1_{-1.9}^{+5.5}$ when evaluating their M and FGK models over the same period and radius range (averaged over $[0.5,256]$\,days, $[0.5,4.0]\,R_\oplus$), but a ratio of $0.9_{-0.2}^{+1.6}$ while evaluating at similar instellations by scaling their FGK grid from a G2 star to match the instellations of an M2.5 star.

\citet{Petigura2022} modeled occurrence distributions in both orbital period and instellation for three bins of stellar mass with boundaries of $\{0.5,0.7,1.0,1.4\}$\,M$_\odot$ using the California-\kepler{} Survey with stellar properties derived from spectroscopic measurements (independent of \gaia{}). Referencing their Figure~(13), their models of sub-Neptune ($1.7-4.0$\,R$_\oplus$) occurrence rates suggest a statistically distinct factor of $\sim3-4$ offset between the lowest and highest bins at the longest periods or lowest instellations, though the underlying separation-binned number of planets per star for each stellar mass bin are not distinct at the $1\sigma$ level. For super-Earths ($1-1.7$\,R$_\oplus$), their models were cut off before reaching the habitable zone due to sparse detections at low instellations, and the underlying binned estimates (including some upper limits) were again not statistically distinct. Combined with \citet{Hsu2020}, these results call into question whether there is sufficient evidence supporting higher occurrence of \textit{Earth-sized} planets at habitable zone instellations around lower-mass stars.

Efforts to resolve potential stellar mass dependencies in the occurrence distribution of small planets face more difficulties when using instellation as opposed to orbital period or semimajor axis. In addition to requiring precise measurements of both stellar mass and luminosity to calculate instellation flux, the true scale factor between occurrence rates around different spectral types may only be resolvable at low instellations (where candidate detections are currently sparse). This is because the occurrence distributions resembling broken power laws appear to ``break" at similar orbital periods (see e.g., \citealp{Petigura2022, Bergsten2022}) across FGKM stars, such that the breaks occur at different values of semimajor axes \citep{Mulders2015} or instellations \citep{Petigura2022}. Instellation exhibits a more prominent offset between spectral types due to a stronger stellar mass dependence when converting from orbital period,\footnote{From Equations~(\ref{eqn:KeplersLaw}) and~(\ref{eqn:Flux}), $a\propto M_*^{1/3}P^{2/3}$ while $I\propto L_* M_*^{-2/3}P^{-4/3}$. Since the mass-luminosity relation $L_* \propto M_*^\xi$ for main sequence dwarfs always has $\xi>1$, the orbital period-instellation conversion has the stronger stellar mass dependence.} meaning that the distribution for higher-mass stars breaks and subsequently ``plateaus" (depending on the power law exponent) at higher instellations where the distribution for lower-mass stars may still be exhibiting a steep rise.

The resulting trend is that higher-mass stars may appear to have slightly higher occurrence rates at high instellations but lower rates at low instellations compared to lower-mass stars (a trend which is not observed when using semimajor axis or orbital period). Only the trend at low instellations is relevant to the habitable zone, and thus only candidates at low instellations can reveal a potential stellar mass dependence. However, the characteristic power law break for low mass stars may fall at instellations where very few relevant candidates have been detected. From \citet{Petigura2022}, plateaus should arise by $\sim10$\,I$_\oplus$ for stars down to $0.5$\,M$_\odot$, while less-massive stars would presumably flatten out at even lower instellation values. Yet \kepler{}'s limited sensitivity to Earth-sized planets orbiting M dwarfs at instellations less than $10$\,I$_\oplus$ (Figure~\ref{fig:modern_samples}) leaves very few detections with which to characterize this regime.

Despite the aforementioned complications, instellation may be the most physically relevant dimension for studies in the context of the habitable zone and/or planet formation. Incident flux from the host star likely contributes to shaping the planet formation environment within a protoplanetary disk -- consider e.g., the disk's temperature profile and the location of the snow line. As cautioned in \citet{Hsu2020}, we note that stellar luminosities are higher on the pre-main sequence than on the main sequence (especially true for M dwarfs), such that the incident flux a planet receives is different between the time of formation and the present day. Additionally, M dwarfs remain on the pre-main sequence at the time of planet formation, such that the relation between disk temperature, instellation and stellar mass differs from pre-main sequence to main sequence. Nevertheless, instellation may still offer a slightly more native tracer of the formation environment compared to semimajor axis or orbital period which include additional stellar mass dependencies. Furthermore, the location of the habitable zone for an Earth-like atmosphere is set by the incident stellar flux \citep{Kopparapu2013}, and thus the habitable zone falls at comparable instellations (but different semimajor axes and orbital periods) for stars of various effective temperatures. This is true both within the M dwarf regime (see Table~\ref{tab:etaEarth}), and across the range of FGKM stars, such that instellation is the most well-suited dimension for comparisons across spectral types.

\subsection{No Evidence for an Increase in $\eta_\oplus$ between M and FGK stars}

\begin{figure*}[htb!]
    \centering
    \includegraphics[width=\textwidth]{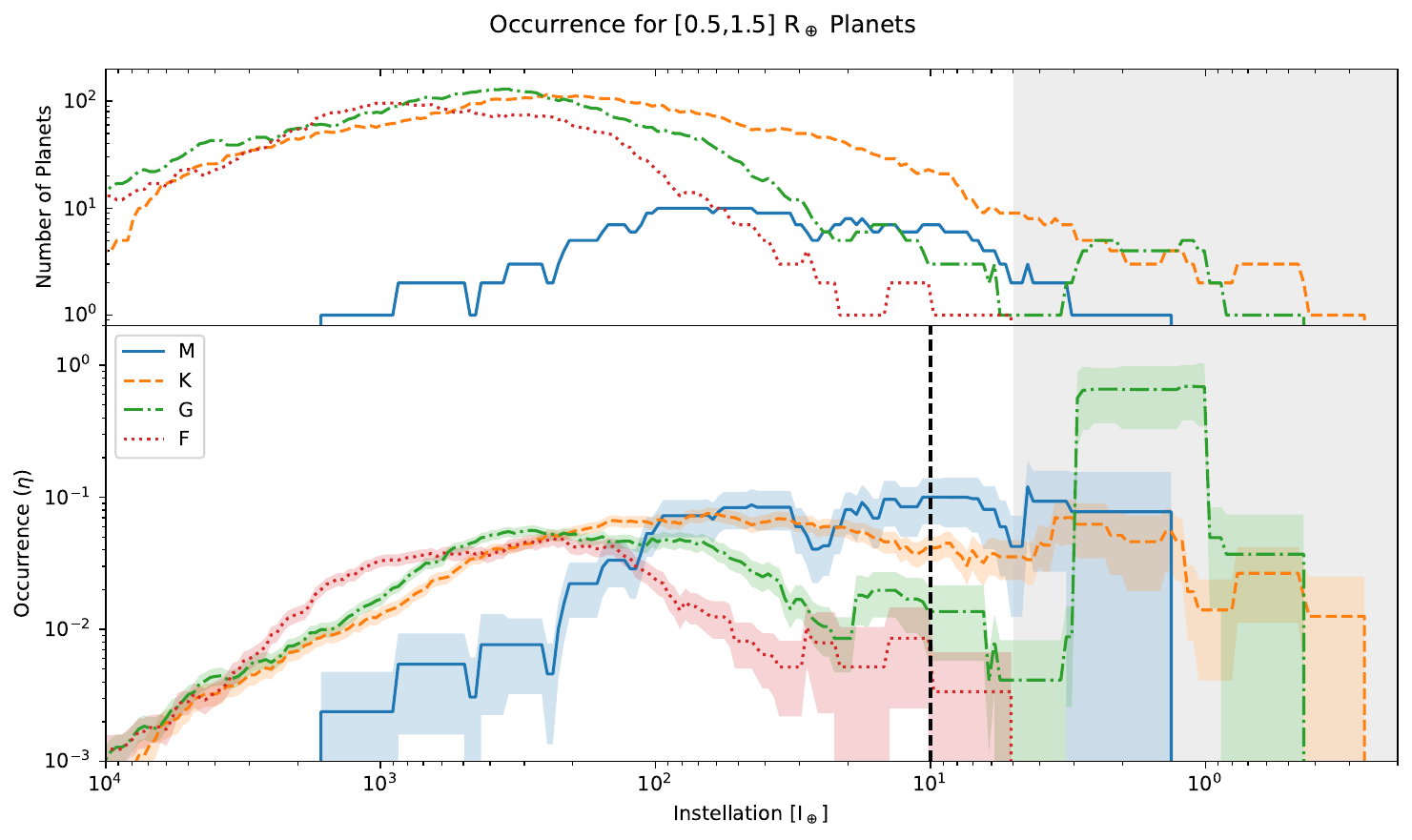}
    \caption{Because of \kepler{}'s poor sensitivity to habitable zone planets (around any star), the current sample does not offer evidence to support M dwarfs hosting more Earth-sized planets than FGK stars \textit{at habitable zone instellations}. \textbf{Top:} The number of Earth-sized ($[0.5,1.5]$\,R$_\oplus$) \kepler{} planet candidates orbiting M (solid blue), K (dashed orange), G (dot-dashed green) or F (dotted red) dwarf stars as a function of instellation. The shaded grey region denotes the regime where there are no detections of Earth-sized planets around F type stars, such that FGKM comparisons require models and/or extrapolations. \textbf{Bottom:} Earth-sized planet occurrence rates calculated via the reliability-weighted inverse detection efficiency method (Equation~\ref{eqn:IDEM}) over a continuum of habitable zone-width instellation bins. Vertical black dashed line represents the approximate point where the \citet{Petigura2022} occurrence distribution plateaus for $[0.5,0.7]$\,M$_\oplus$ stars, indicating the regime where comparisons between spectral types would not be affected by differences in the break points of their power law distributions.}
    \label{fig:FGKM}
\end{figure*}

Before we present a comparison of $\eta_\oplus$ estimates between M and FGK stars, we find it useful to first discuss the current limitations of \kepler{} data regarding Earth-sized ($[0.5,1.5]$\,R$_\oplus$) planets around different spectral types. To this end, we divided the \citet{Berger2020} stellar sample and subsequent planets into M, K, G, and F bins of stellar effective temperatures with bounds of $\{2310,3890,5325,5960,7310\}$\,K following the spectral type definitions of \citet{PecautMamajek, Mamajek}. For each spectral type, we adopted a bin width equivalent to that of the conservative habitable zone for that bin's median stellar effective temperature \citep{Kopparapu2013}. We then slid this bin over a continuum of instellation values, and present the number of detected Earth-sized planets as a function of instellation in the upper panel of Figure~(\ref{fig:FGKM}).

As mentioned in Section~(\ref{sec:intro}), the limited duration of \kepler{}'s observations meant that the survey was generally unable to provide confident detections in the habitable zone for Sun-like stars. The longest orbital period measurable within \kepler{}'s duration while still meeting the three-transit detection requirement is $\sim710$\,days. However, the inner edge of the conservative habitable zone exceeds this limit for $\sim30\%$ of \kepler{}'s F-type stars, such that \kepler{} outright lacked the coverage to detect habitable zone planets around many hotter stars (see e.g., Figure 1 of \citealp{Bryson2021}). For example, there were zero detections of Earth-sized planets around F-type stars at instellations less than $5\,I_\oplus$; more generally, all spectral types have relatively few detections at less than $10\,I_\oplus$.

Because of the paucity of detected Earth-sized planets at low instellations, the \kepler{} sample is inherently ill-suited for direct estimations of occurrence rates in this regime. To illustrate this point, we used the aforementioned sliding bins and the inverse detection efficiency method (Equation~\ref{eqn:IDEM}) to create an occurrence distribution as a function of instellation for each spectral type, shown in the bottom panel of Figure~(\ref{fig:FGKM}). The inverse detection efficiency method is unable to provide occurrence estimates in unpopulated bins, and thus offers the most native visualization of where occurrence estimates are currently motivated by available data.

Compared to the well-sampled close-in regime, occurrence rates at low instellations are more uncertain (and trends therein more inconsistent), especially at the plateau-relevant separations ($<10\,I_\oplus$) necessary to resolve stellar mass dependencies. While there is evidence for increased occurrence around M dwarfs (relative to FGK stars) at $10\,I_\oplus$, it is unclear how this trend persists at habitable zone instellations an order of magnitude lower, due to ``noise" from the generally sparse sample and the lack of relevant detections for F-type stars. As such, we argue that \textbf{the current \kepler{} sample lacks sufficient evidence to inflate $\eta_\oplus$ between FGK and M stars, as the paucity (or lack) of detections for Earth-sized habitable zone planets means that a purely data-motivated comparison remains elusive.}

We thus turn to population models, which offer a way to provide globally informed estimates in these low-detection regimes. However, we caution that model-driven estimates can require assumptions of functional forms and how those forms behave in regions of sparse or absent data. This can obfuscate insight on what is presently motivated by available data, such as the \kepler{} sample's limitations discussed above. With these caveats stated, we now draw comparisons between our $\eta_\oplus$ estimate for M dwarfs and recent literature estimates for Sun-like stars.

Our \texttt{M3} instellation-modeled occurrence rate of $\eta_\oplus={8.58}_{-8.22}^{+17.94}\%$ in this work is similar to the FGK-based value of $\eta_\oplus={9.37}_{-2.48}^{+3.40}\%$ found in \citet{Bergsten2022} which also incorporated reliability and \gaia{}-revised properties, though the former is considerably more uncertain. The latter was modelled in orbital period rather than instellation but used the same definition of the habitable zone, and employed a different range for Earth-sized planets ($[0.7,1.5]$,R$_\oplus$). Normalizing both values by the $\mathrm{d}\ln \mathrm{R}$ of their respective radius ranges still provided consistent results of ${7.81}_{-7.48}^{+16.33}\%$ and ${12.29}_{-3.25}^{+4.46}\%$, respectively. We note that the $\sim3.5$ difference from \citet{Mulders2015b} is still possible beginning at the $\sim1.8\sigma$ level ($\sim$\nth{96} percentile), found by treating both radius-normalized estimates as idealized split-normal distributions and taking their ratio.

The $\eta_\oplus$ estimate of \citet{Bergsten2022} is something of a small-to-intermediate estimate compared to other recent and similarly comprehensive works (see their Section 4.2.1 for details). On the larger side, \citet{Bryson2021} also employed \texttt{DR25} and \gaia{}-revised stellar properties along with treatments of completeness and reliability. They used a different parameterization from that of \citet{Bergsten2022}, opting for a three-dimensional model with dependencies on instellation, planet radius, and stellar effective temperature. For $[0.5,1.5]\,R_\oplus$ planets in a conservative habitable zone around stars with $3900 < T_\mathrm{eff} < 6300$\,K, their $\eta_\oplus$ estimate fell between $37_{-21}^{+48}\%$ and $60_{-36}^{+90}\%$. Some more recent estimates including both reliability and \gaia{}-revised properties include $\eta_\oplus = 11_{-6}^{+7}\%$ from \citet{KunimotoMatthews2020} for $[0.75, 1.5]\,R_\oplus$ planets across $[0.99,1.70]$\,au for G-type stars, and $\eta_\oplus = 12_{-5}^{+8}\%$ from \citet{Bryson2020} for $[0.5,1.5]\,R_\oplus$ planets across $[237,860]$\,days orbiting GK stars. Estimates that include \gaia{}-revised properties but not reliability include $\eta_\oplus = 16_{-6}^{+9}\%$ from \citet{Hsu2019} for $[0.75,1.5]\,R_\oplus$ across $[237,500]$\,days orbiting FGK stars, and $\eta_\oplus = 5_{-4}^{+7}\%$ from \citet{Pascucci2019} for $[0.7,1.5]\,R_\oplus$ planets across $[330,803]$\,days orbiting G-type stars (their Model \#6). Despite the range of recent FGK $\eta_\oplus$ estimates, our estimate of $\eta_\oplus={8.58}_{-8.22}^{+17.94}\%$ for M dwarfs has such large uncertainties that it is consistent with all quoted results within $1\sigma$. To this end, even model estimates do not currently offer justification to inflate $\eta_\oplus$ between FGK and M dwarf stars.

While FGK stars are grouped together in this discussion, it is worth noting that the habitable zone boundaries for FGK stars exhibit a strong dependence on stellar effective temperature (in either instellation or orbital period). This causes the (log) width of the habitable zone to change considerably across the FGK regime \citep{Kopparapu2013}. Corrections for this may include modeling a stellar effective temperature dependence \citep{Bryson2021} or normalizing $\eta_\oplus$ by both $\mathrm{d}\ln \mathrm{P}$ and $\mathrm{d}\ln \mathrm{R}$ (i.e., $\Gamma_\oplus$), although \citet{Bergsten2022} found no statistically significant trend across FGK stars with the latter.


\subsection{Regarding Occurrence Trends within M Dwarf sub-Spectral Types}\label{sec:subMs}

Previous works have suggested that small planet occurrence may exhibit a dependence on \textit{sub}-spectral type across the broad range of M dwarf classifications, though the exact behavior is presently not well-constrained. \citet{KHU2019} studied small planets around \kepler{}'s mid-M dwarfs using revised stellar properties, and found their median occurrence rates to increase from M3 to M5 spectral types (though all values were consistent at $1\sigma$). On the theoretical side, \citet{Mulders2021} predicted that planet occurrence peaks around early ($\sim0.5$\,M$_\odot$, roughly M1) M dwarfs and decreases towards later sub-spectral types.
This decrease in occurrence may explain why recent efforts searching \emph{K2}'s $\sim$M5.5-M9.5 stars \citep{Sagear2020, SestovicDemory2020} or the EDEN survey's volume-limited sample of M7-M9 stars \citep{Dietrich2023} have not identified \textit{any} Earth-sized planets (the latter only studying planets with $P<1$\,day). It is worth noting that late M dwarfs are quite faint and thus difficult to observe with current sensitivity limitations, such that even the Transiting Exoplanet Survey Satellite (\emph{TESS}, \citealp{Ricker2015}) may miss out on detecting tens of transiting planets around nearby late M dwarfs \citep{Brady2022, Dietrich2023}.

\citet{KHU2019} estimated an occurrence rate of $0.99_{_-0.50}^{+0.66}$ for $[0.5,1.5]$\,R$_\oplus$ planets within $[0.5,10]$\,days, calculated using their combined sample of $13$ \kepler{} planets around seven mid-M dwarfs. We evaluated our \texttt{M3} orbital period model across the same range to yield a statistically distinct estimate of $0.26^{+0.07}_{-0.06}$. The inconsistency between estimates could potentially be attributed to an overestimation caused by their simplified treatment of detection efficiency, and/or the very small sample size in \citet{KHU2019} leading to a non-robust measurement.

The mid-M dwarf regime is also bright enough to be studied by \emph{TESS}, offering additional insight beyond \kepler{}. \citet{Ment2023} used \emph{TESS} to study a sample of mid-to-late M dwarfs (M4 to M7) with a median stellar mass of $0.17$\,M$_\odot$, and identified a sample of seven planets. Spanning $[0.5,7]$\,days orbital period, they found an occurrence rate of $0.13_{-0.07}^{+0.12}$ for $[0.5,1.0]$\,R$_\oplus$ planets and $0.45_{-0.14}^{+0.19}$ for $[1.0,1.5]$\,R$_\oplus$. Our sample (including all dwarfs below $4000$\,K) has a median mass of $0.50$M$\odot$ such that it is dominated by earlier M dwarfs, and we found $0.09_{-0.03}^{+0.04}$ and $0.10_{-0.02}^{+0.03}$ respectively for the same period and radius ranges using our \texttt{M3} models in orbital period. Despite the different spectral types probed by these studies, our results are consistent with \citet{Ment2023} in the $[0.5,1.0]$\,R$_\oplus$ bin. Our estimate is considerably smaller for $[1.0,1.5]$\,R$_\oplus$ planets, which could be attributed to our use of a single power law to describe the radius distribution, while \citet{Ment2023} used a normal distribution that peaks in the $[1.0,1.5]$\,R$_\oplus$ bin.

However, while comparing the same radius bins evaluated over $[4,200]$\,I$_\oplus$ in instellation, \citet{Ment2023} reported $0.11_{-0.06}^{+0.10}$ and $0.37_{-0.12}^{+0.16}$ while we found $0.18_{-0.06}^{+0.08}$ and $0.21_{-0.05}^{+0.06}$. Our results are consistent with theirs at the $1\sigma$ level in either radius bin, potentially suggesting that \citet{Hsu2020}'s finding of FGK/M stars having comparable occurrence at similar instellations also applies to an early/mid-to-late M dwarf comparison \textit{for Earth-sized planets}. The size stipulation is critical: because our modeling efforts focused solely on Earth-sized planets, we cannot offer insight to \citet{Ment2023}'s comparison of $[0.5,4]$\,R$_\oplus$ planet occurrence rates between their mid-to-late sample and \citetalias{DressingCharbonneau2015} representing early Ms,\footnote{The \citetalias{DressingCharbonneau2015} sample was originally believed to have a median stellar effective temperature of $3746$\,K (roughly M0.5; \citealp{Mamajek}). Our \gaia{}-informed updates to the \citetalias{DressingCharbonneau2015} sample described in Section~(\ref{sec:DC15}) suggest an actual median temperature of $4096$\,K (roughly K7).} nor the results of \citet{KHU2019} for mid-Ms which included planets up to $2.5$\,R$_\oplus$. While incorporating larger planets up to $4$\,R$_\oplus$ would enable more detailed comparisons, this would likely require more complicated functional forms to properly address the full radius distribution or coupled dependencies, and is thus beyond the scope of this work.

Our focus on a relatively small radius regime left a limited number of detected planets which, combined with our top-heavy sample of host star temperatures (Figure~\ref{fig:stars}), meant our study was unable to resolve any relationship between sub-spectral type and the occurrence of Earth-sized planets using \kepler{}. Details of our attempt to probe such a dependence and limitations therein are discussed further in Appendix~(\ref{app:TeffSplit}). To address these shortcomings, we discuss the potential of future studies employing data from \kepler{} and additional surveys in the following subsection.

\subsection{Improving M Dwarf Occurrence Rates with Additional Surveys}\label{sec:future}
The \emph{K2} \citep{Howell2014} and \emph{TESS} \citep{Ricker2015} missions will likely help to further constrain $\eta_\oplus$ around M dwarfs by providing more observations of Earth-sized habitable zone planets. Similar to the \gaia{}-\kepler{} Stellar Properties Catalog of \citet{Berger2020}, \citet{KHU2020} provided a uniform catalog of \gaia{}-informed stellar properties for the \emph{K2} population. The \emph{K2} planet candidate sample of \citet{Zink2021} also contains $47$ $[0.5,1.5]\,R_\oplus$ planets around $T_\mathrm{eff} < 4000\,K$ stars, including two Earth-sized candidates: a $0.51\,R_\oplus$ planet in the conservative habitable zone, and a $1.20\,R_\oplus$ planet in the optimistic (see their Table 1). \citet{Zink2023} already combined the \kepler{} and \emph{K2} samples to study close-in planets around FGK stars, and similar integrated studies in the future may be well-poised to (re)address occurrence comparisons between FGK/M stars, and explore occurrence trends between early, mid, and late M dwarfs as discussed in Section~(\ref{sec:subMs}).

A single \emph{TESS} sector is observed for $\sim27$\,days, which (requiring two transits for detectability) probes the habitable zones up to M4.5 dwarf stars. The Continuous Viewing Zone at the ecliptic poles covers a smaller area of sky and thus views less stars, but it is observed for $\sim350$\,days which provides a long enough baseline to survey the habitable zone even for M0 stars. While later spectral types offer habitable zones at shorter orbital periods, they also exhibit increased stellar activity \citep{Robertson2013, West2015, AstudilloDefru2017}, and the resulting increase in photometric noise may contribute to \emph{TESS}'s unexpectedly low yield of candidate detections around later M dwarfs \citep{Brady2022}. Nonetheless, the study of \emph{TESS}'s mid-to-late M dwarfs (M4 to M7) from \citet{Ment2023} estimated a $\sim50\%$ sensitivity to $1$\,R$_\oplus$ planets at $7$\,days (or $4$\,I$_\oplus$; their Figure~6), such that there may be appreciably non-zero sensitivity at the slightly larger habitable zone separations needed to probe $\eta_\oplus$ and subsequent stellar mass dependence. Future survey missions such as \emph{PLATO} \citep{Rauer2014, Rauer2016} and the \emph{Nancy Grace Roman Space Telescope} \citep{Spergel2015, Akeson2019} may also help to constrain $\eta_\oplus$ by enriching the sample of small planets around M dwarfs.

On a more contemplative note, future studies may be able to leverage abundant planet detections to adopt a more physical definition of an ``Earth-sized" planet beyond a generic $[0.5,1.5]\,R_\oplus$ bin. For example, \citet{KHU2023} adopted an instellation-dependent lower bound set by the minimum planet mass capable of retaining an Earth-like atmosphere \citep{ZahnleCatling2017, BixelApai2021}, and an upper bound where a planet orbiting an M dwarf is likely to have some particularly hospitable atmospheric composition \citep{KimuraIkoma2022}. An alternative upper bound could be the limit beyond which small planets are no longer rocky in composition, conservatively interpreted as $1.4$\,R$_\oplus$ \citep{Rogers2015} as used in the LUVOIR \citep{LUVOIR} and HabEx \citep{HabEx} mission concept studies. These may be considered in addition to coupled size-instellation boundaries like the planet mass-dependent inner edge from \citet{Kopparapu2014} used in this work (see Appendix~\ref{app:mass-dependent-HZ}), and/or treatment of atmospheric factors such as clouds that may vary between Earth- and super-Earth-sized worlds (e.g., \citealp{Windsor2023}). If/when model uncertainties are reduced enough such that these considerations would have meaningful effects on $\eta_\oplus$ estimates, it may be worth building towards a more physically meaningful set of integral bounds for habitable zone occurrence rates.

\section{Summary}\label{sec:Conclusions}
Since the work of \citet{DressingCharbonneau2015}, \gaia{} updates to stellar properties have changed our understanding of \kepler{}'s M dwarfs and their sample of planet candidates, including the Earth-sized habitable zone planets used to estimate $\eta_\oplus$. Here, we presented an updated investigation of the current \kepler{} sample, fitting separate population models in orbital period and instellation with various considerations of reliability and parameter uncertainties.

\begin{itemize}
    \item Using \kepler{} \texttt{DR25}, \gaia{}-updated parameters and candidate reliability, we found that the updated \kepler{} sample has few detected candidate planets in the habitable zone. As such, the inverse detection efficiency method cannot be employed as in the past to calculate $\eta_\oplus$.
    
    \item Integrating our best-fit instellation model incorporating both reliability and parameter uncertainties, we estimated an occurrence rate for Earth-sized habitable zone planets of $\eta_\oplus={8.58}_{-8.22}^{+17.94}\%$ for the conservative habitable zone (and ${14.22}_{-12.71}^{+24.96}\%$ for the optimistic). Our orbital period model predicted a slightly larger median value of ${13.89}_{-9.02}^{+16.70}\%$ (${21.75}_{-13.23}^{+23.44}\%$) which is consistent within the $1\sigma$ uncertainties.

    \item The updated sample's paucity of Earth-sized habitable zone planet detections means that \kepler{} offers no evidence supporting an increased $\eta_\oplus$ around M dwarfs compared to FGK stars. This also applies to model estimates in the literature, as our $\eta_\oplus$ value for M dwarfs is consistent with those based on FGK stars.
\end{itemize}

We note that, in either dimension, the uncertainties on our $\eta_\oplus$ estimates are typically larger than previous works. This is largely due to the small number of \kepler{} Earth-sized candidates at larger orbital periods or lower instellations, requiring us to rely on extrapolated models whose parameters suffer similar uncertainties from the paucity of detections. The median values are also generally lower than previous works, which may be (at least partially) attributed to our consideration of candidate reliability. Compared to our model(s) in orbital period, our instellation model is less well-defined (Figure~\ref{fig:breaks}) and provides occurrence estimates with larger uncertainties. Yet instellation offers a more native tracer of the planet formation environment and processes defining the habitable zone, and is thus our preferred dimension for population models and comparisons across spectral types.

When evaluated in instellation, we found a lack of sufficient evidence that would support M dwarfs having more Earth-sized planets than FGK stars at habitable zone instellations. This contrasts with studies of close-in planets evaluated in orbital period and/or semimajor axes, leaving open the possibility that scaling habitable zone occurrence rates with spectral type may not be justified.
We also did not find a significant difference between occurrence rates for our predominantly early M dwarf sample and those of mid-to-late M dwarfs, though we could only compare estimates for Earth-sized planets where our models are defined. Future studies with \emph{K2} and \emph{TESS} should be able to further probe spectral type dependencies, especially in the habitable zones of M dwarfs.


\section{Acknowledgements}
I.P. and G.B acknowledge support from the NASA Astrophysics Data Analysis Program under Grant No. 80NSSC20K0446.
K.H-U. and R.B.F. acknowledge support from NASA under Agreement No. 80NSSC21K0593 for the program ``Alien Earths”. G.D.M. acknowledges support from FONDECYT project 11221206, from ANID --- Millennium Science Initiative --- ICN12\_009, and the ANID BASAL project FB210003.
This research has made use of the NASA Exoplanet Archive, which is operated by the California Institute of Technology, under contract with the National Aeronautics and Space Administration under the Exoplanet Exploration Program.
The results reported herein benefited from collaborations and/or information exchange within NASA’s Nexus for Exoplanet System Science (NExSS) research coordination network sponsored by NASA’s Science Mission Directorate.

\vspace{5mm}
\facilities{\gaia{}, \kepler{}}

\software{\texttt{NumPy} \citep{numpy}, \texttt{SciPy} \citep{scipy}, \texttt{Matplotlib} \citep{pyplot}, \texttt{emcee} \citep{ForemanMackey2012}, \texttt{corner} \citep{corner}, \texttt{epos} \citep{Mulders2018}, \texttt{KeplerPORTs} \citep{BurkeCatanzarite2017}}

\appendix

\section{False Positive Dispositions}\label{app:FPs}

A catalog of homogeneously characterized objects is essential to the uniform analysis required for demographic studies. \kepler{}'s \texttt{DR25} satisfies this requirement by virtue of its automation, but we nevertheless acknowledge that some objects automatically classified as false positives in \texttt{DR25} may have received new classifications from subsequent works. We briefly discuss the sample of relevant objects here, but stress that any updated dispositions do not come from uniform treatments applied to the entire planet sample and are thus not used in our demographic study.

The Exoplanet Archive's table of \kepler{} Certified False Positives summarizes the efforts of the \kepler{} False Positive Working Group to revisit and more rigorously classify potential false positives in the \kepler{} sample \citep{Bryson2017}. Of the five  \citet{DressingCharbonneau2015} planets labeled false positives in \texttt{DR25}, one was a certified false positive and four were "potential planets," although one star was not included in \citet{Berger2020} and two others received updated $T_\mathrm{eff} > 4000$\,K. The one remaining planet, K00961.02 (Kepler-42 c), is listed as a confirmed planet in the Exoplanet Archive. In our updated investigation based on the full \texttt{DR25} sample (Section~\ref{sec:ModernSample} onward), there were nine \texttt{DR25} false positives that orbited an M dwarf (via \citealp{Berger2020}) and had planet parameters within our sample domain (in orbital period, instellation, or both). Other than the aforementioned K00961.02, six of these were certified false positives, and the remaining two were "possible planets," although one had a reliability score of $\sim1\%$ and is thus unlikely to be real. Again the remaining planet, K03138.02 (Kepler-1649 c), is listed as a confirmed planet in the Exoplanet Archive. Even though K03138.02 has a habitable zone instellation of $0.94 \pm 0.09\,I_\oplus$, we reiterate that its disposition status in the literature cannot supersede its \texttt{DR25} classification by virtue of the uniformity requirement necessitated for demographic study.

\section{Discussing Numerical Choices Related to Model Optimization}\label{app:mcmc}
All numbers of total and discarded steps in this work were chosen based on autocorrelation analysis with \texttt{emcee}. In general, we required the total chain length to surpass at least $50$ times the autocorrelation time $\tau$, and discard at least $2\tau$ as burn-in. For optimization with the \texttt{M2} and \texttt{M3} approaches, the number of iterations was the same as those used in \citet{Bryson2021}: 100 draws when including reliability, and 400 when including both reliability and input uncertainties. While we lack a quantitative justification, we found that 100 inferences suitably incorporates candidates at a rate linearly proportional to their reliability. Similarly, 400 inferences was a suitable number of iterations such that unreliable candidates could be included in enough draws to also sample the range of their uncertainties.

\section{Investigating the Bimodality in Instellation}\label{app:bimodality}

Here, we investigated the cause of the instellation bimodality shown in Figure~(\ref{fig:breaks}), and discuss whether this feature can be attributed to two physically distinct populations or an intrinsic scatter from our relatively sparse sample. Figure~(\ref{fig:modern_samples}) shows a relative paucity of $[1.0,1.5]\,R_\oplus$ planet detections around $10\,I_\oplus$ separating two clusters of observed planets, roughly corresponding to the trough between the two peaks in the $\log_{10} I_\mathrm{break}$ posteriors of Figure~(\ref{fig:breaks}). This gap in detections manifests itself as a slight decrease in occurrence around $10\,I_\oplus$ (see estimates from the inverse detection efficiency method in Figure~\ref{fig:model_F}) -- such that a broken power law could place the break (a transition between rising and $\sim$plateauing occurrence) on either side of this dip -- but this variation is not statistically significant.

\begin{figure}
    \centering
    \includegraphics[width=\textwidth]{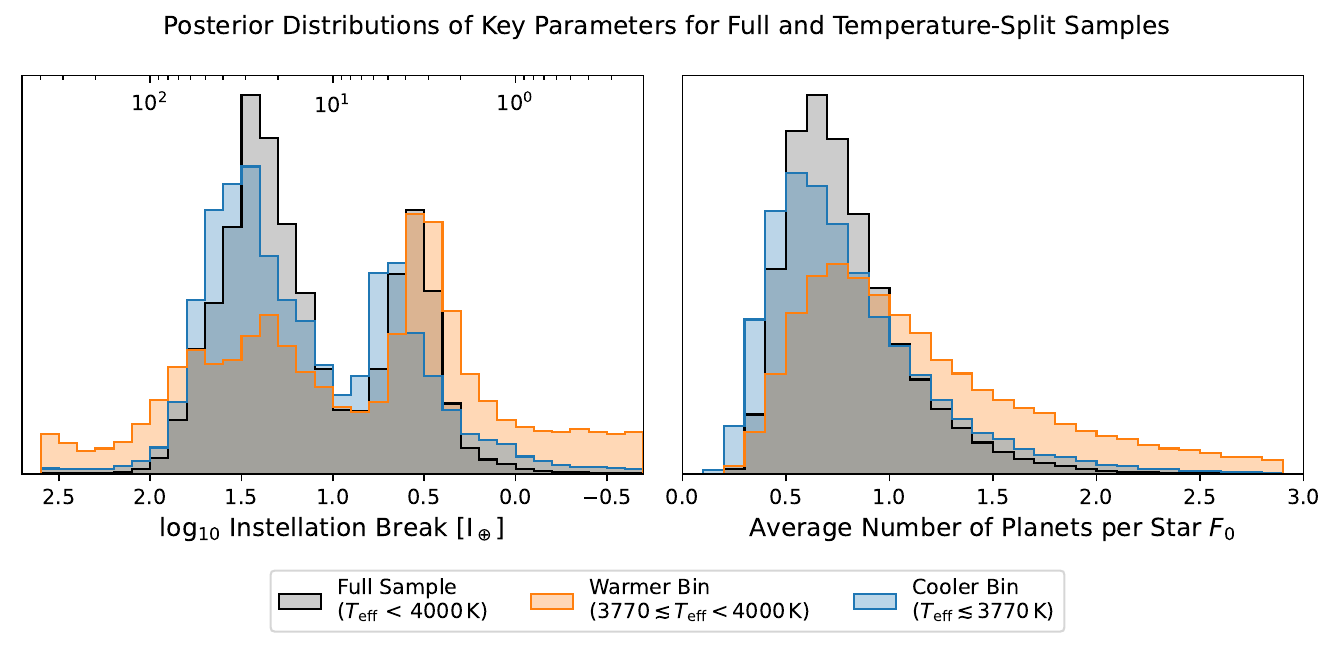}
    \caption{Posterior distributions for the \textbf{left:} $\mathrm{log}_{10}$ power law breaks in instellation and \textbf{right:} average number of planets per star. Grey histograms represent posteriors from the \texttt{M3} models for the full sample (same as Figure~\ref{fig:breaks}), while the orange and blue histograms show posteriors from models fit to the warmer and cooler subsamples split at $T_\mathrm{eff}\sim3770$\,K). In the left panel, tick marks on the upper edge represent the corresponding non-logarithmic values. Both temperature bins produce a bimodality in $\log_{10} I_\mathrm{break}$, suggesting this feature is not attributable to two host star populations distinct in temperature, though the location of the primary peaks is opposite what one would expect from scaling the same orbital period to hotter/cooler stars. In the right panel, we cannot resolve any significant difference in the average number of Earth-sized planets between warmer and cooler M dwarfs.}
    \label{fig:flux_breaks}
\end{figure}

We explored a potential physical explanation for the instellation bimodality. Given the unimodal distribution of $\log_{10} P_\mathrm{break}$ posteriors, if the power law break occurs at the same orbital period for all M dwarfs, then this break could manifest at two different instellations for two distinct enough groups of host stars (distinguished on the basis of stellar mass, luminosity and effective temperature). To test if this was the case, we split the M dwarf sample at the median stellar effective temperature ($\sim3770$\,K) to produce two bins with an equal number of stars, and repeat the \texttt{M3} fitting procedure for 100 inferences in either bin. The resulting models' instellation break posterior distributions are plotted in the left panel of Figure~(\ref{fig:flux_breaks}) alongside the full sample posteriors from Figure~(\ref{fig:breaks}). Neither of the bins had unimodal $\log_{10} I_\mathrm{break}$ distributions, and both bins shared peaks at $\log_{10} I_\mathrm{break} \approx 0.6$ and $1.4$, or $I_\mathrm{break} \approx 4$ and $25\,I_\oplus$ respectively. The cooler bin's primary (stronger) peak occurs at the latter, while the warmer bin's primary peak occurs at the former.

This contradicts an expectation that, if the power law break occurs at the same orbital period for all stars, then the break should correspond to smaller instellation values for less massive (and thus cooler/fainter) stars. Nonetheless, because both the warmer and cooler subsamples exhibit the same general bimodality, we found it unlikely that the effective temperature distribution of host stars is responsible for this feature in the $\log_{10} I_\mathrm{break}$ posterior distribution. Lacking alternative physical explanations, we could not assign a root cause for the bimodality, and considering such possibilities would (a) likely require a larger sample of planets more thoroughly sampling host star parameter space, and (b) be  beyond the scope of this work. Given our present limitations, we attributed the instellation break bimodality to noise within our small sample which manifests as two possibilities for $\log_{10} I_\mathrm{break}$ under our functional form of choice.

\subsection{An Additional Outcome of the Temperature-Split Model}\label{app:TeffSplit}

In the right panel of Figure~(\ref{fig:flux_breaks}), we show the posterior distributions for the scaling parameter $F_0$ representing the average number of planets per star from models fit to the full, warmer, and cooler subsamples. We found a value of $F_0 = 1.05_{-0.42}^{+0.92}$ for the warmer subsample (median temperature of $\sim3894$\,K), and a value of  $F_0 = 0.72_{-0.26}^{+0.45}$ for the cooler subsample (median temperature of $\sim3614$\,K), which are fully consistent at the $1\sigma$ level. We note that the accompanying uncertainties are large enough that the warmer/cooler ratio could fall anywhere from $\sim0.5 - 4.3$ within $1\sigma$.

As such, with these models we were not able to make any claims regarding the dependence of $F_0$ on sub-spectral type discussed in Section~(\ref{sec:subMs}). This is most likely due to our very limited sample of relevant stars and planets. Focusing on only Earth-sized planets already limits the available sample of candidates to which one would fit a population model, even more so when splitting the sample up further based on host stellar effective temperature. While it may be true that \kepler{} lacks the candidates to resolve a finer stellar dependence for \textit{Earth-sized} planets around M dwarfs, a less restrictive (though more involved) study considering all small planets ($<4$\,R$_\oplus$) may have enough candidates to probe such trends. However, the distribution of host star temperatures for those small planets is still top-heavy, making it difficult to explore mid and late M dwarfs with \kepler{}. As noted in Sections~(\ref{sec:subMs}) and~(\ref{sec:future}), additional surveys with improved sensitivity to such targets could provide valuable information contribution to future studies, especially those capable of integrating information from multiple surveys for greater coverage and resolution.

\section{Planetary Mass Dependence of Habitable Zone Bounds}\label{app:mass-dependent-HZ}

Here we describe our approach to adopting the planet-mass-dependent scaling of \citet{Kopparapu2014} to modulate the conservative inner (i.e., runaway greenhouse) edge with planet mass. We used the three example planetary masses ($0.1,1.0$ and $5.0\,M_\oplus$) in \citet{Kopparapu2014} to calculate the corresponding runaway greenhouse edges for each star in our sample. We then took the average bound from all stars for each example planet mass, and used these to define a one-dimensional interpolation to estimate the runaway greenhouse flux for a given planet mass. Because our models were defined in planet radius while the above is defined in planet mass, when integrating our models across the habitable zone in planet radius and instellation, we employed an empirical mass-radius relation \citep{ChenKipping2017} to translate each radius point to a corresponding mass, then used the above interpolation to compute the corresponding flux defining the inner edge of the conservative habitable zone at that radius. In \citet{Kopparapu2014} the outer conservative edge (maximum greenhouse) and the optimistic edges were not believed to change with planet mass, so these occur at constant fluxes with respect to planet size in our integrations.

\bibliography{ref}{}
\bibliographystyle{aasjournal}

\end{document}